\documentclass[%
 reprint,
superscriptaddress,
 amsmath,amssymb,
 aps
]{revtex4-2}
\pdfoutput=1

\usepackage{graphicx}% Include figure files
\usepackage{dcolumn}% Align table columns on decimal point
\usepackage{bm}% bold math
\usepackage{amsmath,mathrsfs,amssymb,dsfont}
\usepackage{xcolor}
\usepackage{import}
\usepackage{enumitem}
\usepackage[normalem]{ulem}
\usepackage{soul}
\usepackage{physics}
\usepackage{comment}
\usepackage{float}
\usepackage{afterpage}
\usepackage{pdfpages}

\makeatletter
\AtBeginDocument{\let\LS@rot\@undefined}
\makeatother

\begin{document}

\preprint{APS/123-QED}

\title{Nanometer-Scale Nuclear Magnetic Resonance Diffraction with Sub-\AA ngstrom Precision}

\author{Holger Haas}
%\altaffiliation{H. Haas and S. Tabatabaei contributed equally to this work.}
\altaffiliation[Present address: ]{IBM Quantum, IBM T.J. Watson Research Center, Yorktown Heights, NY 10598, USA}
\affiliation{Department of Physics and Astronomy, University of Waterloo, Waterloo, ON, Canada, N2L3G1}
\affiliation{Institute for Quantum Computing, University of Waterloo, Waterloo, ON, Canada, N2L3G1}

\author{Sahand Tabatabaei}
%\altaffiliation{H. Haas and S. Tabatabaei contributed equally to this work.}
\affiliation{Department of Physics and Astronomy, University of Waterloo, Waterloo, ON, Canada, N2L3G1}
\affiliation{Institute for Quantum Computing, University of Waterloo, Waterloo, ON, Canada, N2L3G1}

\author{William Rose}
\affiliation{Department of Physics, University of Illinois at Urbana-Champaign, Urbana, Illinois 61801, USA}

\author{Pardis Sahafi}
\affiliation{Department of Physics and Astronomy, University of Waterloo, Waterloo, ON, Canada, N2L3G1}
\affiliation{Institute for Quantum Computing, University of Waterloo, Waterloo, ON, Canada, N2L3G1}

\author{Mich\`ele Piscitelli}
\altaffiliation[Present address: ]{Clarendon Laboratory, Department of Physics, University of Oxford, OX1 3PU, UK}
\affiliation{Department of Physics and Astronomy, University of Waterloo, Waterloo, ON, Canada, N2L3G1}
\affiliation{Institute for Quantum Computing, University of Waterloo, Waterloo, ON, Canada, N2L3G1}

\author{Andrew Jordan}
\affiliation{Department of Physics and Astronomy, University of Waterloo, Waterloo, ON, Canada, N2L3G1}
\affiliation{Institute for Quantum Computing, University of Waterloo, Waterloo, ON, Canada, N2L3G1}

\author{Pritam Priyadarsi}
\affiliation{Department of Physics and Astronomy, University of Waterloo, Waterloo, ON, Canada, N2L3G1}
\affiliation{Institute for Quantum Computing, University of Waterloo, Waterloo, ON, Canada, N2L3G1}

\author{Namanish Singh}
\affiliation{Department of Physics and Astronomy, University of Waterloo, Waterloo, ON, Canada, N2L3G1}
\affiliation{Institute for Quantum Computing, University of Waterloo, Waterloo, ON, Canada, N2L3G1}

\author{Ben Yager}
\altaffiliation[Present address: ]{Oxford Instruments NanoScience, Abingdon, Oxfordshire, OX13 5QX, UK}
\affiliation{Department of Physics and Astronomy, University of Waterloo, Waterloo, ON, Canada, N2L3G1}
\affiliation{Institute for Quantum Computing, University of Waterloo, Waterloo, ON, Canada, N2L3G1}

\author{Philip J. Poole}
\affiliation{National Research Council of Canada, Ottawa, Ontario, Canada, K1A 0R6}

\author{Dan Dalacu}
\affiliation{National Research Council of Canada, Ottawa, Ontario, Canada, K1A 0R6}

\author{Raffi Budakian}
\email{rbudakian@uwaterloo.ca}
\affiliation{Department of Physics and Astronomy, University of Waterloo, Waterloo, ON, Canada, N2L3G1}
\affiliation{Institute for Quantum Computing, University of Waterloo, Waterloo, ON, Canada, N2L3G1}

\date{\today}% It is always \today, today,

%%% Abstract %%%
\begin{abstract}
Achieving atomic resolution is the ultimate limit of magnetic resonance imaging (MRI), and attaining this capability offers enormous technological and scientific opportunities, from drug development to understanding the dynamics in interacting quantum systems.
In this work, we present a new approach to nanoMRI utilizing nuclear magnetic resonance diffraction (NMRd) --- a method that extends NMR imaging to probe the structure of periodic spin systems.
The realization of NMRd on the atomic scale would create a powerful new methodology for materials characterization utilizing the spectroscopic capabilities of NMR.
We describe two experiments that realize NMRd measurement of $^{31}$P spins in an indium-phosphide (InP) nanowire with sub-\AA ngstrom precision. In the first experiment, we encode a nanometer-scale spatial modulation of the $z$-axis magnetization by periodically inverting the $^{31}$P spins, and detect the period and position of the modulation with a precision of $<0.8~\text{\AA}$. In the second experiment, we demonstrate an interferometric technique, utilizing NMRd, for detecting an \AA ngstrom-scale displacement of the InP sample with a precision of 0.07\,\text{\AA}. The diffraction-based techniques developed in this work represent new measurement modalities in NMR for probing the structure and dynamics of spins on sub-\AA ngstrom length scales, and demonstrate the feasibility of crystallographic MRI measurements.
\end{abstract}

\maketitle
\section{Introduction}
Three-dimensional atomic-resolution nuclear magnetic resonance (NMR) imaging of nanometer scale materials remains a long-standing challenge. The ability to image biologically-relevant structures such as proteins and virus particles on the atomic scale with the spectroscopic capabilities of NMR would fundamentally advance our understanding of their function, and potentially lead to new drug therapies \cite{Ajoy2015,Perunicic2016,Schirhagl2014}.
In addition, atomic-scale magnetic resonance imaging (MRI) could be a powerful tool for studying magnetic correlations in condensed matter systems \cite{Casola2018}.
Over the past two decades, significant advances have been made to extend the capabilities of MRI to the nanometer and \AA ngstrom scales \cite{Rugar2004,Grinolds2013,Willke2019,Lovchinsky2016,Degen2009,Taminiau2012,Zopes2018,Grob2019,Rugar2015,Arai2015,Ziem2019,Rose2018}, including the detection of single sub-surface electron spins \cite{Rugar2004,Grinolds2013}, MRI of single atoms on a surface \cite{Willke2019}, NMR detection of single proteins \cite{Lovchinsky2016}, three-dimensional imaging of individual tobacco-mosaic virus particles \cite{Degen2009}, and the detection and coherent control of individual carbon spins in diamond \cite{Taminiau2012} with sub-\AA ngstrom resolution \cite{Zopes2018}. 

Many of these approaches rely on the use of local probes to directly detect the distribution of spins within the sample. While such techniques are well-suited for imaging arbitrary spin distributions, scattering techniques that employ coherent sources, e.g., X-rays, neutrons or electrons, offer an alternative means of determining the atomic-scale structure of crystalline materials that possess a high degree of spatial correlation. Unlike direct-space techniques that locally probe the material density, scattering techniques rely on the interference of the scattered field from each scattering center within the detection volume, providing a highly efficient means of structure determination.

MRI, like X-ray and neutron scattering, is a reciprocal space technique, in which the measured signal $s(\mathbf{k})\sim\int\rho(\mathbf{r})\,e^{i\mathbf{k}\cdot\mathbf{r}}\,d^3r$ is proportional to the Fourier transform of the matter density.
This similarity with scattering led Mansfield and Grannell in 1973 to propose NMR `diffraction' (NMRd), as a method for determining the lattice structure of crystalline materials \cite{Mansfield1973}. 
The realization of crystallographic NMRd would extend the spectroscopic capabilities of MRI to the atomic scale and provide a fundamentally new non-destructive method for measuring the structure factor of crystalline materials.
Furthermore, being a phase-sensitive technique, NMRd would permit real-space reconstruction of the spin density, without the loss  of phase information common to scattering techniques, such as X-rays, that measure the scattered field intensity.

The main challenge to achieving atomic scale NMRd lies in the difficulty of making the encoding wavenumber $k$ sufficiently large \footnote{In MRI, $k=\gamma G \tau$ is the encoding wavenumber corresponding to the spatial modulation of the nuclear spin magnetization, generated by the evolution of spins with gyromagnetic ratio $\gamma$ in the magnetic field gradient $G$ for a time $\tau$.}. For example, the largest encoding wavenumbers achieved in clinical high-resolution MRI scanners are of order $k/(2\pi)\sim 10^4\,\text{m}^{-1}$, which is more than a factor of $10^5$ smaller than what is needed to measure typical atomic spacings \cite{Vachha2021}.
In the past two decades, the principal technologies needed to overcome this challenge have been developed in the context of force-detected nanoMRI, paving the way for realizing atomic-scale NMRd, which would establish a new modality for materials characterization and imaging.

In this work, we present two experiments that utilize key advances in nanoMRI technology --- namely the ability to generate large time-dependent magnetic field gradients, and the ability to detect and coherently control nanoscale ensembles of nuclear spins --- to generate encoding wavenumbers as large as $k/(2\pi)=0.48\ \text{\AA}^{-1}$ and realize NMRd measurements of $^{31}$P spins of an indium phosphide (InP) sample with sub-\AA ngstrom precision. These results represent a significant step towards extending the spectroscopic and phase-sensitive imaging capabilities of MRI to atomic-scale materials characterization.

In the first experiment, we demonstrate phase-sensitive NMRd detection by encoding a `diffraction grating' via periodic modulation of the $^{31}$P $z$-axis magnetization, with a mean period of 4.5~nm in a $(\sim50~\text{nm})^3$ volume, and detect the position and period of the grating with a precision of $<0.8~\text{\AA}$. In the second experiment, we present a method for interferometric displacement detection using NMRd, which we apply to measure an \AA ngstrom-scale displacement of the InP sample with a precision of 0.07\,\text{\AA}.

\section{NMR\MakeLowercase{d} Concept}

\begin{figure}[th]
    \centering
    \includegraphics[width = \columnwidth]{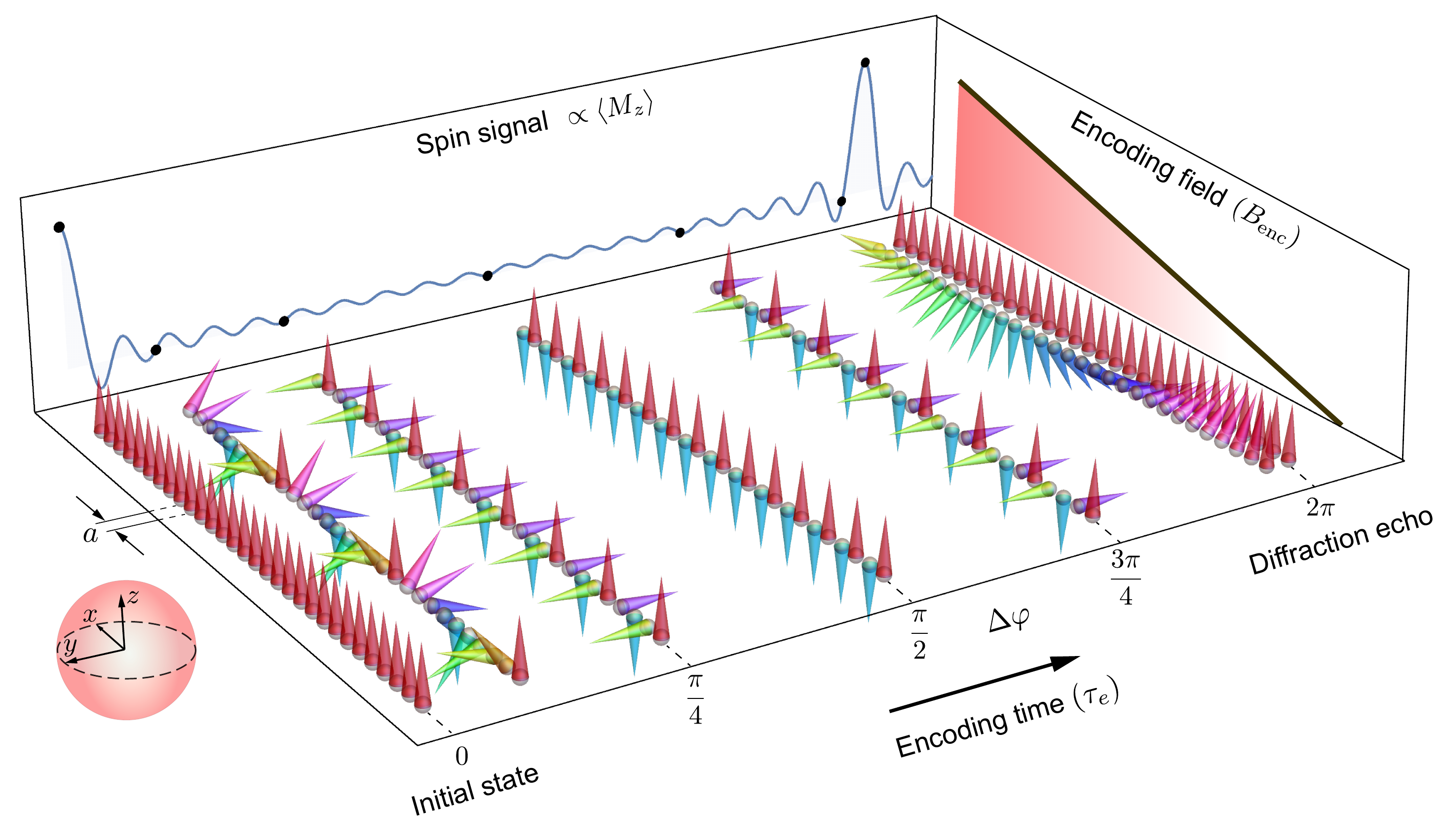}
    \caption{Time evolution of a one-dimensional periodic lattice of spins, starting from a uniform $z$~state at $\tau_e=0$, under a linearly varying external field along the lattice. The $x, y, z$ coordinate axes on the left marks the Bloch sphere directions for the spins represented by the cones. For $\tau_e>0$, the spins dephase and the expectation value of the $z$-axis magnetization $\expval{M_z}$ drops to zero. Once the spins on adjacent lattice sites complete a full rotation $(\Delta\varphi=2\pi)$, the spins rephase and a diffraction echo in $\expval{M_z}$ appears.}
    \label{fig:Fig1}
\end{figure}

To illustrate the basic concept of NMRd as envisioned in Ref.~\cite{Mansfield1973}, we consider a one-dimensional spin density having a spatially periodic modulation---such as a linear spin chain with spacing $a$ as shown in Fig.~\ref{fig:Fig1}---that evolves in a uniform field gradient $G_x=\partial B/\partial x$ for a time $\tau_e$. % HH: Added previously missing white-spaces
The wavevector corresponding to the helical winding encoded in the spins is $k_x=\gamma\,G_x\,\tau_e$, where $\gamma$ is the spin gyromagnetic ratio.
At particular encoding times $\tau_{\text{echo}}=2\pi n/(\gamma\,G_x\,a)$, corresponding to $k_x a=2\pi n$, the relative phase between neighboring spins becomes $\Delta\varphi=2\pi n$, $n\in\{1,2,3,\dots\}$ and a `diffraction echo' (DE) is observed.
At the peak of the echo, the signal from each spin adds constructively, in exact analogy to the diffraction peak observed in a scattering experiment. The lattice constant is determined from the location of the DE peak, and the shape of the sample from the Fourier transform of the DE envelope. Because the encoding wavevector is spin selective, the structure factor corresponding to each NMR-active nucleus can be determined separately.
The NMRd concept illustrated in Fig.~\ref{fig:Fig1} can be readily generalized to three dimensions, with $\mathbf{k}=\gamma\mathbf{G}\tau_e$ and $\mathbf{G}=\grad B$, where $B$ represents the magnitude of either the static or the radio frequency (RF) field at the Larmor frequency, used for phase encoding.
The diffraction condition in three dimensions corresponds to $\mathbf{k}\cdot \mathbf{a}_j = 2\pi n$, where $\mathbf{a}_j$ are the primitive vectors of the lattice.

Being particularly sensitive to hydrogen atoms, NMRd could enable structural characterization of nanocrystalline organic solids via NMR. For example, a lattice of $^1$H spins with $a=3~\text{\AA}$ evolving under a uniform field gradient of $10^5$~T/m would produce a DE at $\tau_e=780~\mu\text{s}$. While the dephasing times in most organic solids are much shorter, typically of order $T_2\sim 20~\mu\text{s}$, dynamical decoupling NMR pulse sequences, such as the symmetric magic echo sequence \cite{Boutis2003}, can be used to extend the coherence time into the millisecond range, while allowing for encoding with both static and resonant RF field gradient pulses.
Importantly, although the concept of NMRd was first envisioned as a technique to study crystal structures, it can be applied more broadly to probe any spatially-periodic spin-state modulation, e.g., a periodic modulation of the $z$-axis magnetization, that can be refocused by the evolution under the field gradient.
It can therefore also be used to study quantum transport of periodic spin systems on atomic length scales.
\section{Experimental Setup}
\begin{figure}[h]
    \centering
    \includegraphics[width = \columnwidth]{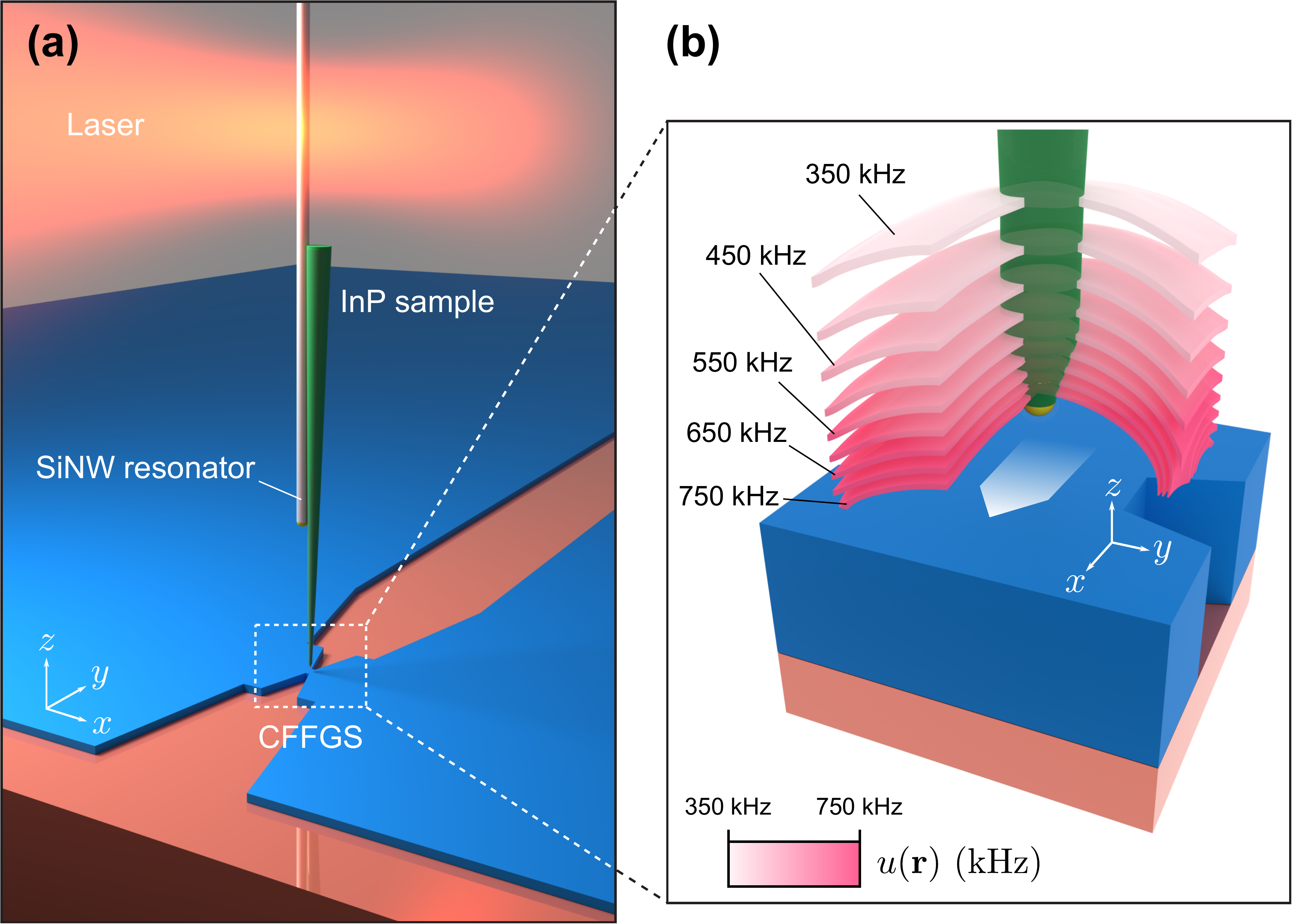}
    \caption{(a) Schematic of the experimental setup, including the SiNW force sensor, the current-focusing field-gradient source (CFFGS), and the InP spin sample.
    During spin detection, 70-mA peak current was applied to the CFFGS at the resonance frequency of the SiNW, which produced a time-dependent field gradient with a peak amplitude of $\partial B_z/\partial y =$~8~G/nm, 50~nm above the surface. The Rabi field $B_1(\mathbf{r})=\sqrt{B_x^2(\mathbf{r})+B_y^2(\mathbf{r})}/2$ used for NMR was produced by driving the CFFGS with 70~mA-pk currents at the resonance frequency of  $^{31}$P spins, which produced a $B_1$ field with an amplitude of 470~G, and a $B_1$ field gradient of $\partial B_1/\partial z=-4$~G/nm, 50~nm above the surface. Further details regarding the field profile produced by the CFFGS are provided in Sec. II of Ref.~\cite{Supplements}. (b) Simulated contours of constant Rabi frequency $u(\mathbf{r})=\gamma B_1(\mathbf{r}) / (2\pi)$, where $\gamma=2\pi\times 17.235$~MHz/T is the $^{31}$P gyromagnetic ratio. The contours within the sample are nearly parallel to the $xy$~plane and vary primarily in the $z$ direction.}
    \label{fig:Fig2}
\end{figure}
Force-detected magnetic resonance measurements were performed using a silicon nanowire (SiNW) mechanical resonator, which served as the mechanical sensor to detect the force exerted on $^{31}$P spins in an InP nanowire (InPNW) placed in a magnetic field gradient. The experimental setup, shown in Fig.~\ref{fig:Fig2}(a), is similar to the one used in our previous nanoMRI work \cite{Rose2018}.
The SiNW was grown via the vapor-liquid-solid method near the edge of a $1.5~\text{mm}\times1~\text{mm}\times0.4~\text{mm}$ Si(111) substrate, and had a length of 20~$\mu$m, and a 100~nm diameter \cite{Sahafi2020}.
The frequency of the fundamental flexural mode of the as-grown SiNW was approximately 250~kHz prior to sample attachment, and had a spring constant of 0.6~mN/m.
Experiments were carried out at a base temperature of 4~K in high vacuum.
At this temperature, the quality factor of the SiNW was approximately 60,000. The SiNW chip was glued to a millimeter size piezoelectric transducer (PT), which was used to apply various control signals to the SiNW. To increase the measurement bandwidth of the resonator, a feedback signal was applied to the PT, which reduced the quality factor of the SiNW to 700 \cite{Poggio2007}.

The InPNW sample used in this work was approximately 5-$\mu$m long, with $\sim 100$-nm diameter, grown with a Wurtzite structure \cite{Dalacu2012}.
The sample was attached $\sim3-4$~$\mu$m away from the tip of the SiNW, with the axes of the two nanowires aligned parallel to each other, and the tip of the InPNW extending $\sim 2\,\mu\text{m}$ beyond the tip of the SiNW.
Details of the attachment procedure are provided in Sec.~I of Ref.~\cite{Supplements}.
A video of the sample attachment is also included in \cite{Supplements}.
After sample attachment, the resonance frequency of the SiNW decreased to 163~kHz, however no significant change in the quality factor was observed.

NMR measurements were performed by applying a static field of $B_0=3$~T parallel to the the InPNW axis. At this field, the Larmor frequency of the $^{31}$P spins is $\omega_0/(2\pi)=51.7\,\text{MHz}$. To generate time-varying magnetic fields and magnetic field gradients used for spin measurements, we fabricated a current focusing field gradient source (CFFGS) by electron-beam lithography and reactive ion-beam etching of a 100-nm thick Al film deposited on a sapphire substrate. The device contained a 150-nm-wide and 50-nm-long constriction, which served to focus electrical currents to produce the magnetic fields used for spin detection and control. All measurements were carried out with the tip of the InPNW placed at the center of the CFFGS and positioned $\sim50$~nm above the surface.

\section{Nanometer-Scale NMR\MakeLowercase{d} Measurements}\label{NMRd_18X_grating}
\begin{figure}[h]
    \centering
    \includegraphics[width = \columnwidth]{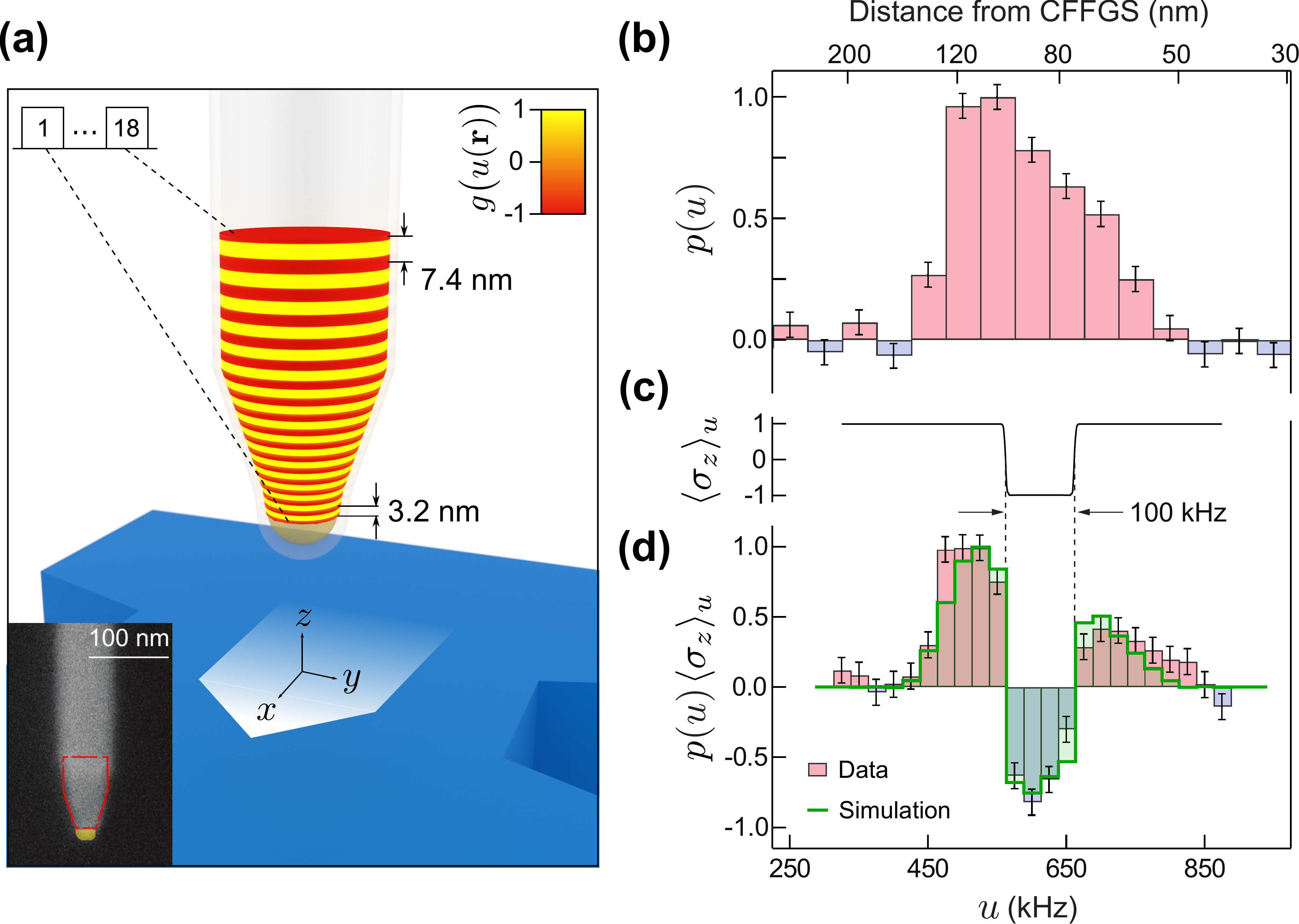}
    \caption{(a) Spatial configuration of the InP spin sample brought 50~nm above the surface of the CFFGS device. The CFFGS surface lies perpendicular to the static external magnetic field $B_0~\hat{\mathbf{z}}$.
    The 18 regions indicated in red represent the regions inverted by the band-inversion pulses within $g(u)$.  (Inset) Scanning electron microscope (SEM) image of a representative InPNW. The measured volume of the sample is indicated by the outlined region.
    (b) Measured Rabi-frequency distribution $p(u)$ of the $^{31}$P spins in the detection volume shown in (a). The data was obtained by sampling 20 points in the frequency range $0\leq u \leq 1~\text{MHz}$. The upper horizontal axis indicates the $z$~coordinate corresponding to the particular $u$ value on the InPNW axis at the center of the CFFGS $(x=y=0)$.
    (c) $\langle \sigma_z \rangle_u$ profile for the 100~kHz-wide band-inversion pulse, targeting spins in the Rabi frequency range 562.5~kHz $\leq u \leq$ 662.5~kHz. (d) Rabi-frequency distribution measured after applying the band-inversion pulse shown in (c).
    The data was obtained by sampling 23 points in the frequency range $325~\text{kHz} \leq u \leq 875~\text{kHz}$.
    The solid green line indicates the expected distribution calculated using the measured $p(u)$ of (b).}
    \label{fig:Fig3}
\end{figure}

To observe a focused diffraction echo -- i.e., one in which the spectral weight of the echo is localized within a narrow range of encoding times -- the spin-state modulation must be a periodic function of the encoding field coordinate, e.g., for a spin density with a spatially periodic modulation, the encoding field profile must vary linearly in space (Fig. \ref{fig:Fig1}).
As a demonstration of nanometer scale NMRd, we utilize the Rabi-field gradient to (1) create a diffraction grating by periodically inverting the $z$-axis magnetization of the $^{31}$P spins within the measured volume of the InP tip [Fig.~\ref{fig:Fig3}(a)], and (2) generate the encoding wavevector for the NMRd measurements. In so doing, we ensure that the spin modulation is a periodic function of the encoding field.

\begin{figure*}[t]
    \includegraphics[width=\textwidth]{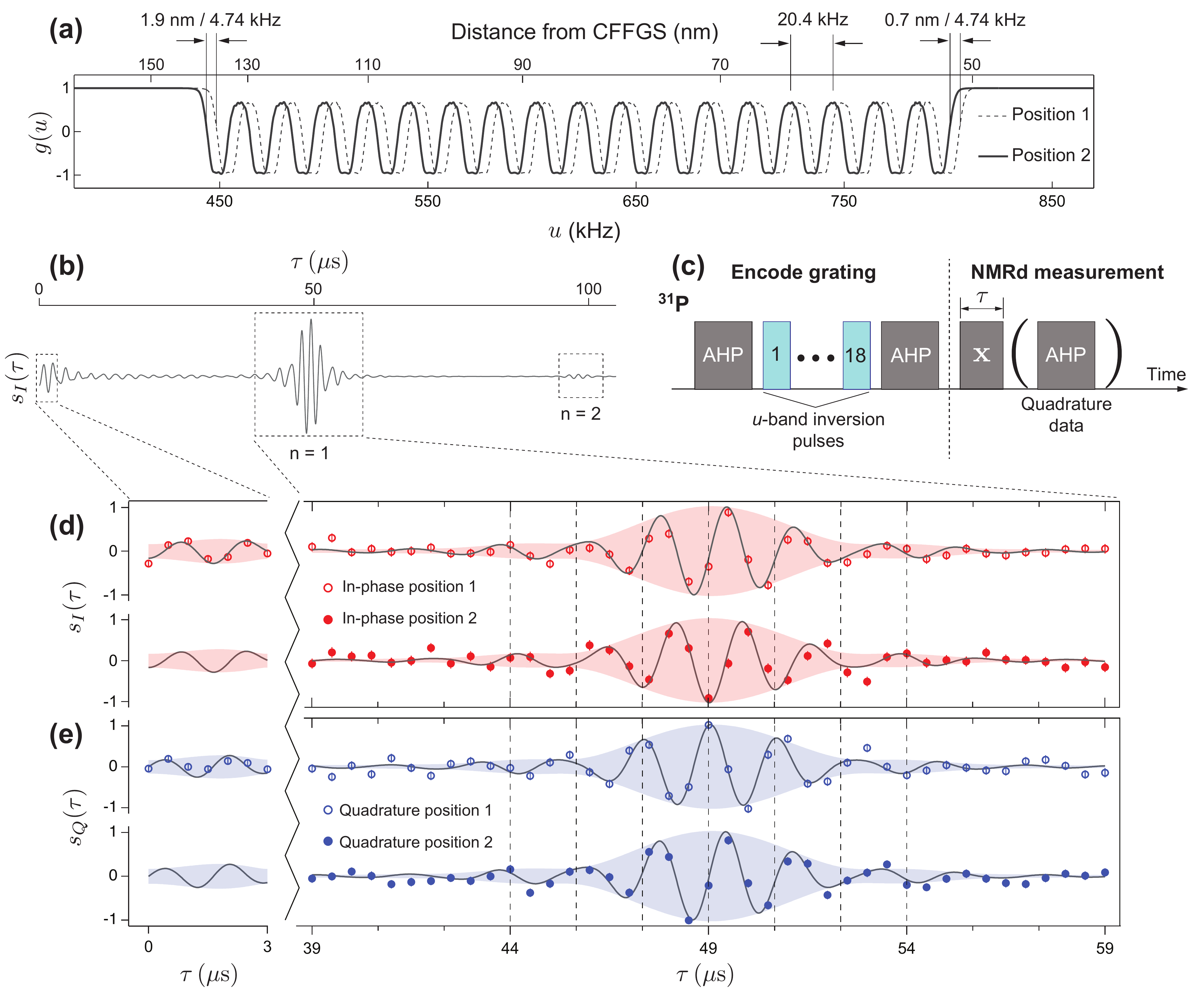}
    \caption{NMRd data measured for the $18\times$ diffraction grating. (a) Inversion profile for the two positions ($g_1$ and $g_2$) of the periodic grating encoded with a period of 20.4~kHz.
    The top horizontal axis indicates the position-dependence of $g_1$ and $g_2$ with respect to the CFFGS. The shift in position corresponding to $\Delta u = 4.74$~kHz is indicated for the regions of the sample that are 50~nm and 140~nm away from the CFFGS.
    (b) Calculated in-phase component $s_I(\tau)$ for the grating at position 1 given the simulated $g_1(u)$ profile in (a) and measured $p(u)$ shown in Fig.~\ref{fig:Fig3}(b).
    (c) NMRd pulse sequence: the encoding part of the sequence encodes either $g_1(u)$ or $g_2(u)$ in (a) by applying 18 consecutive inversions. The second part of the sequence performs the NMRd readout.
    $s_I(\tau)$ is measured by applying a resonant RF pulse $(\mathbf{x})$ for a a duration $\tau$.
    An extra adiabatic half-passage pulse is applied for detecting the quadrature component $s_Q(\tau)$.
    (d, e) $s_I(\tau)$ and $s_Q(\tau)$ measurements for the two grating positions in (a) as a function of the effective encoding time $\tau$.
    The shaded regions in (d, e) indicate the signal amplitude.}
    \label{fig:Fig4}
\end{figure*}

We detect the statistical spin fluctuations in an ensemble of approximately $2\times 10^6$ $^{31}$P spins within the conical region of the InP sample indicated in Fig.~\ref{fig:Fig3}(a) (see Sec.~III of Ref.~\cite{Supplements}), using the MAGGIC spin detection protocol described in Ref.~\cite{Rose2018}.
The measured signal is proportional to the integrated $z$-axis magnetization $s\propto\int\,du\,p(u)\expval{\sigma_z}_u$ (see Appendix~\ref{Appendix}), where $\expval{\sigma_z}_u$ is the expectation value of the Pauli $z$ operator for a spin at Rabi frequency $u$ determined by the NMR encoding sequence used in the MAGGIC protocol, and $p(u)$ is the Rabi-frequency distribution (see Appendix~\ref{Appendix}), determined both by the geometry of the sample as well as the detection protocol which constrains the effective measurement volume near the CFFGS.
We experimentally determine $p(u)$ [Fig.~\ref{fig:Fig3}(b)] using the Fourier encoding method presented in \cite{Rose2018}.

We encode the periodic grating $g(u) = \expval{\sigma_z}_u$ by selectively inverting sequential regions in the sample that are 10.2-kHz wide and separated by 10.2~kHz in $u$, as indicated by Fig.~\ref{fig:Fig4}(a).
The physical regions targeted by the inversions are shown in
Fig.~\ref{fig:Fig3}(a).
To generate a particular $g(u)$, we implement control waveforms that invert spins within the range $(u_{\text{low}}, u_{\text{high}})$ with adjustable edge sharpness around $u_{\text{low}}$ and $u_{\text{high}}$.
To verify the performance of the inversion waveform, we implement a control sequence, shown in Fig.~\ref{fig:Fig3}(c), that targets spins within a 100-kHz bandwidth: $562.5\,\text{kHz} \leq u \leq  662.5\,\text{kHz}$. The measured Rabi-frequency distribution after the application of the band-inversion pulse [Fig.~\ref{fig:Fig3}(d)] agrees closely with the expected inversion profile.
Further details regarding the band-inversion pulses are in Sec.~IV(B) of Ref.~\cite{Supplements}, which also includes an animation depicting the operation of the 100 kHz-wide band-inversion pulse.

The NMRd protocol used to measure $g(u)$ is shown schematically in Fig.~\ref{fig:Fig4}(c).
After encoding the grating, we apply a Larmor-frequency RF pulse for a duration $\tau$.
The evolution of the spin starting from the state $\rho_0 \propto \sigma_z$, when driven with constant Rabi frequency $u$ is described by a unitary $U(u,\tau)=\exp(-i2\pi u \tau\sigma_x/2)$ in the frame rotating at $\omega_0$.
The in-phase $(s_I)$ and quadrature $(s_Q)$ components of diffraction signal are the ensemble-averaged expectation values of $\sigma_z$ and $\sigma_y$:
\begin{align}
    \begin{bmatrix}
        s_I(\tau)\\
        s_Q(\tau)
    \end{bmatrix}&=
    \int_0^\infty\,du\,p(u)\,g(u)
    \begin{bmatrix}
        \mathrm{Tr}\big[\sigma_z U(u,\tau)\rho_0 U^{\dagger}(u,\tau)\big]\\[5pt]
        \mathrm{Tr}\big[\sigma_y U(u,\tau)\rho_0 U^{\dagger}(u,\tau)\big]
    \end{bmatrix}\nonumber\\[5mm]
    &\propto \int_0^\infty\,du\,p(u)\,g(u)
    \begin{bmatrix}
        \cos(2\pi u\tau)\\
        \sin(2\pi u\tau)
    \end{bmatrix}.
    \label{eq1}
\end{align}
To measure the quadrature part of NMRd signal, we end the measurement sequence with a numerically-optimized adiabatic half-passage (AHP) pulse that rotates $\sigma_y$ to $\sigma_z$. It can be seen that if $g(u)$ has a single modulation period $\Omega$, i.e., $g(u+n\Omega)=g(u)$, $n\in\mathbb{Z}$, and varies much more rapidly than the envelope of $p(u)$, then $s_I(\tau)$ and
$s_Q(\tau)$ contain a series of diffraction echos separated by $1/\Omega$-long intervals. The amplitudes of these echos reflect the magnitude of the Fourier coefficients of $g(u)$, and the echo envelopes contain the Fourier transform of $p(u)$.

\begin{figure}[b]
    \centering
    \includegraphics[width=\columnwidth]{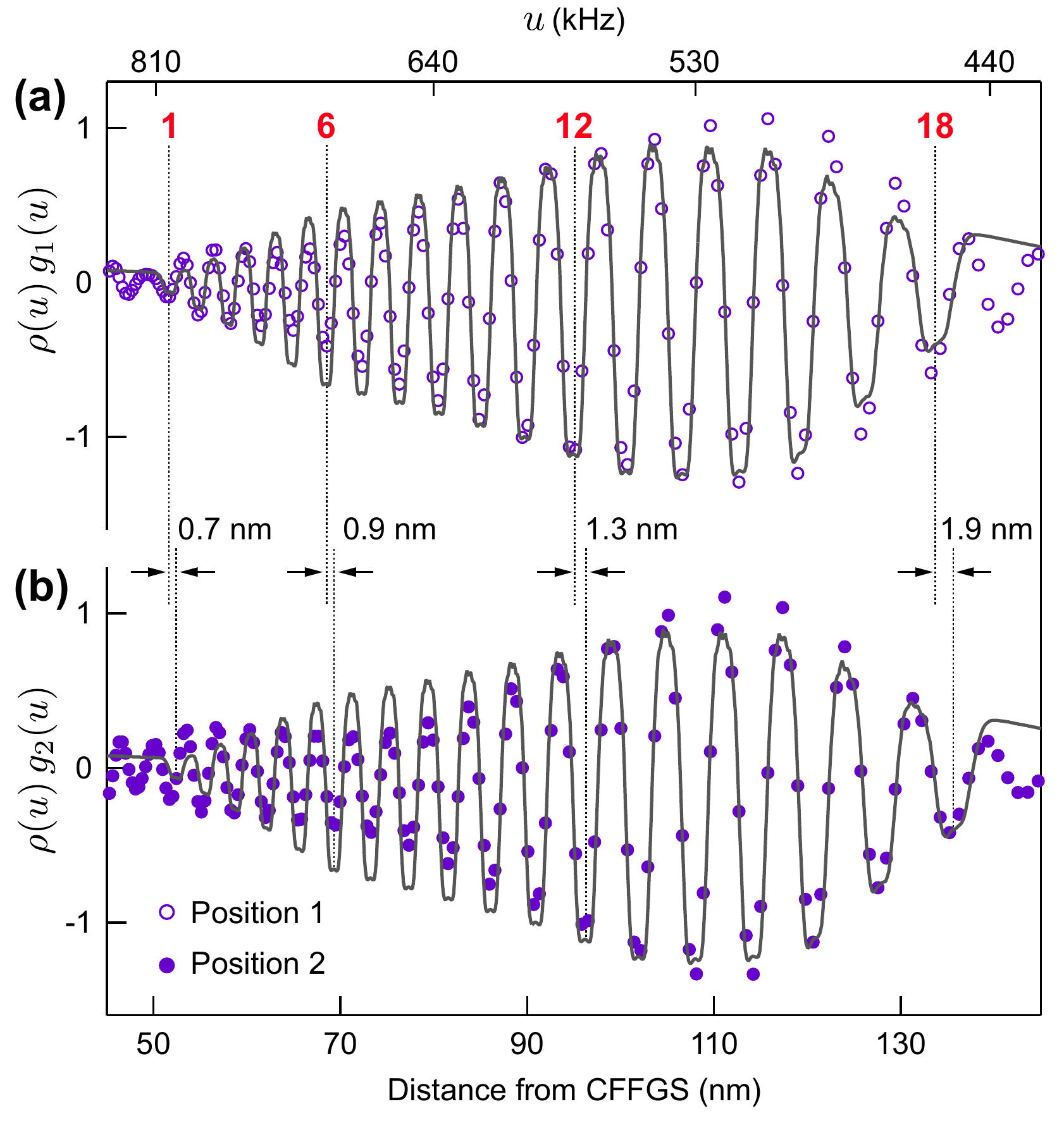}
    \caption{Coordinate-space reconstruction of the diffraction grating.
    Solid lines in (a) and (b) are the calculated $p(u)g_1(u)$ $[p(u)g_2(u)]$ using $g_1(u)$ $[g_2(u)]$ in Fig.~\ref{fig:Fig4}(a) and $p(u)$ in Fig.~\ref{fig:Fig3}(b). The solid dots are the coordinate-space reconstructions of $p(u)g_1(u)$ and $p(u)g_2(u)$ by Fourier transforming the data in Fig. \ref{fig:Fig4} after padding it with zeroes. The position values indicated on the lower horizontal axis are determined from the simulated field distribution produced by the CFFGS (see Sec.~II of Ref.~\cite{Supplements}).The dashed vertical lines are placed as a guide to indicate the spatial offset between the two grating positions. The number of the inversion slice is indicated above the vertical lines.}
    \label{fig:Fig5}
\end{figure}

To demonstrate the phase sensitivity of NMRd, we
encode two spin magnetization gratings $g_1(u)$ and $g_2(u)$, shown in Fig.~\ref{fig:Fig4}(a), that differ by a 4.74~kHz translation, i.e., $g_2(u) = g_1(u + 4.74~\text{kHz})$. We refer to $g_1(u)$ and $g_2(u)$ as the grating at position 1 and 2, respectively. The physical displacement of the grating corresponding to $\Delta u = 4.74~\text{kHz}$ is $\Delta z=0.7$~nm ($\Delta z=1.9$~nm) at $z = 50$~nm ($z = 140$~nm). Both gratings are produced using the control sequences shown in Fig.~\ref{fig:Fig4}(c) that comprise 18 band inversion waveforms sandwiched between two AHPs identical to the ones used to measure $s_Q(\tau)$.

Fig.~\ref{fig:Fig4}(b) shows a plot of the expected $s_I(\tau)$ calculated using Eq.~\eqref{eq1}, $g_1(u)$, and the measured $p(u)$ shown in Fig.~\ref{fig:Fig3}(b). Because the first DE at $\tau\sim 49~\mu\text{s}$
has a significantly larger amplitude than the higher diffraction orders, we measure the NMRd signal only for $n=1$, for both grating positions, in the interval $39~\mu\text{s}\leq\tau\leq 59~\mu\text{s}$, using the NMRd measurement protocol is shown in Fig.~\ref{fig:Fig4}(c).
We note that for all measurements shown in Fig.~\ref{fig:Fig4}, the maximum duration of the resonant RF pulse used in the NMRd measurement portion of the sequence was $59\,\mu\text{s}$, which is much shorter than the transverse relaxation time  $T_{2\rho}=570\,\mu\text{s}$ measured under a continuous resonant RF drive (see Sec.~V of Ref.~\cite{Supplements}). Therefore, we ignore decoherence effects during the Rabi pulse in our analysis and simulations. The resulting data are shown in Fig.~\ref{fig:Fig4}(d, e), which also include the un-scaled calculated values for $s_I(\tau)$ and $s_Q(\tau)$.
The echo envelopes for both signal quadratures are shown as the shaded regions in Fig.~\ref{fig:Fig4}(d,~e). We see that although the small shift in the grating position produces little discernible change in the echo envelope, it is clearly visible in the change in the relative phase of the in-phase and quadrature measurements, demonstrating the importance of phase-sensitive detection for high-precision position measurements.

In Sec.~VI(A, B) of Ref.~\cite{Supplements}, we construct a statistical estimator \cite{Casella_1990} to determine the periodicity $\Omega$ from the measured data. The calculation is done assuming no prior knowledge of $g(u)$, other than the fact that it is periodic in $u$, and varies more rapidly than $p(u)$.
The resulting estimates for $\Omega$ in $g_1$ and $g_2$ are $\Omega_1 = 20.30 \pm 0.14$~kHz and $\Omega_2 = 20.26 \pm 0.10$~kHz, respectively. The 0.14~kHz (0.10~kHz) error in the period of $g_1$ ($g_2$) in $u$-space corresponds to an uncertainty of 0.2~\AA~(0.15~\AA) in the wavelength of the grating at $z = 50$~nm and 0.6~\AA~(0.4~\AA) at $z = 140$~nm.

In Sec.~VI(C, D) of Ref.~\cite{Supplements}, we construct a maximum likelihood estimator for $\Delta u$ for the data in Fig.\ref{fig:Fig4}, which yields $\Delta u = 4.67\pm 0.20$~kHz, in excellent agreement with the expected value of $\Delta u=4.74\,\text{kHz}$. The error in $\Delta u$ corresponds to an uncertainty of 0.3~\AA~(0.8~\AA) in the relative $z-$axis position of $g_1$ ($g_2$) at $z = 50$~nm ($z = 140$~nm).

We reconstruct the periodic spin modulation in coordinate space by calculating the real part of the complex Fourier transform $\Re\{\mathcal{F}^{-1}[s_I(\tau)+i\,s_Q(\tau)]\}$ using the data shown in Fig.~\ref{fig:Fig4}(d, e) for the two grating positions. In the reconstruction, we include the points around the $n=1$ DE, as well as the points sampled near $\tau=0$ to account for a small DC offset in the modulation envelope caused by a slight asymmetry in the magnitude of the positive and negative amplitude regions in $g(u)$. The time records used in the Fourier transforms are constructed to be continuous by zero-padding the un-sampled regions in the interval $0\leq\tau\leq 59.5~\mu\text{s}$ [see Fig. \ref{fig:Fig4}(b)]. The data in the interval $0\leq\tau\leq 3\,\mu\text{s}$ is not expected to change for the two grating positions. Therefore, this interval was measured only for position~1 and used in the reconstruction of both grating positions. 
We see that the position-space representation of $p(u)g_1(u)$ and $p(u)g_2(u)$, shown in Fig.~\ref{fig:Fig5}, closely follow the calculated values.

\section{Displacement Detection via NMR Interferometry}\label{NMR_Interferometer}
\begin{figure*}[!htbp]
    \centering
    \includegraphics[width=\textwidth]{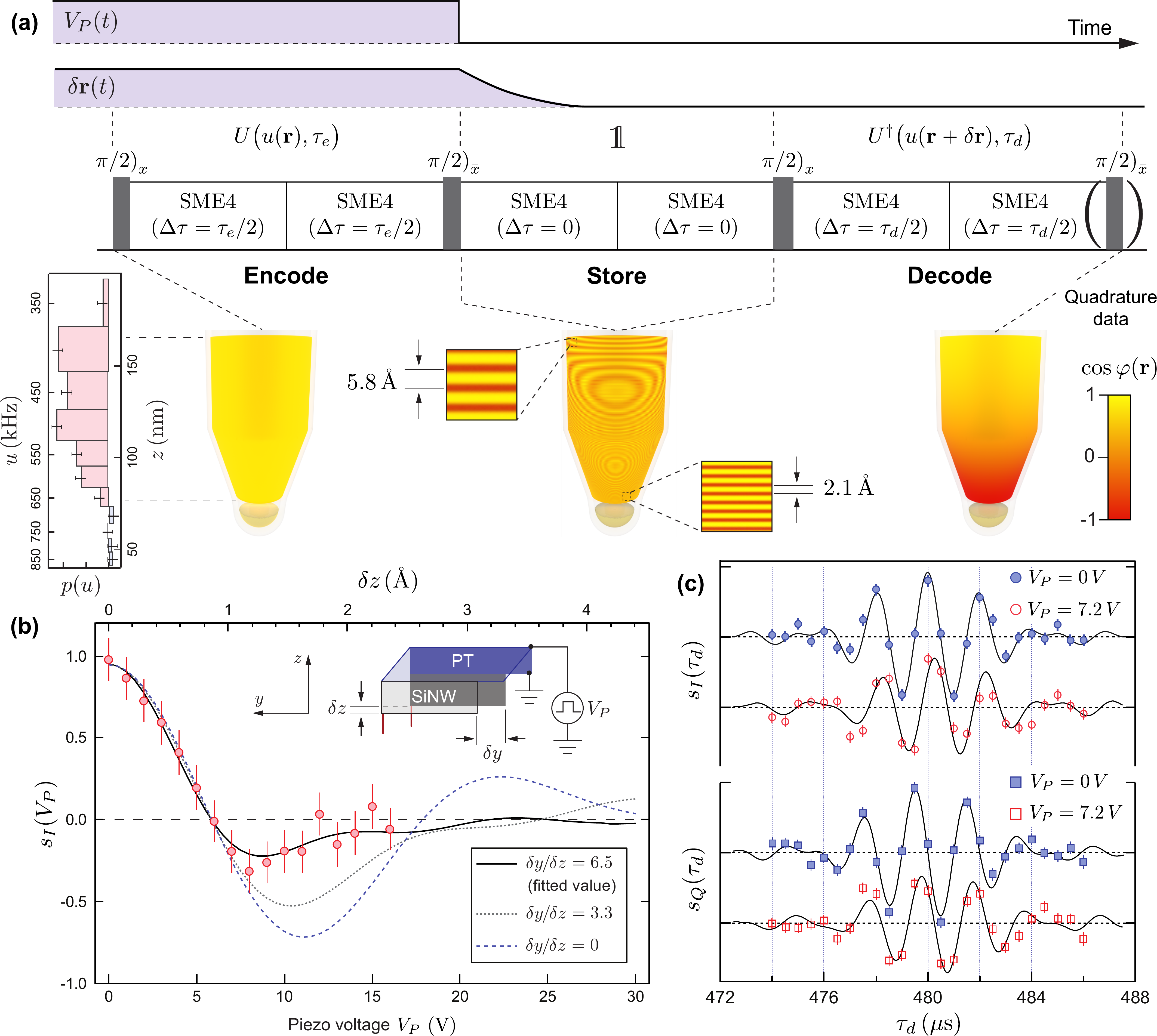}
    \caption{(a) Pulse sequence used for interferometric displacement detection, as well as the the voltage $V_P(t)$ applied to the PT, and the PT displacement $\delta\mathbf{r}(t)$. The unitary operations corresponding to different parts of the sequence are indicated above each block.
    The data presented in (b, c) were acquired from spins in the Rabi frequency range $400\,\text{kHz}<u<650$\,\text{kHz}, corresponding to the sample volume indicated by the shaded regions in (a). The shading indicates the $z$-axis modulation $[\cos\varphi(\mathbf{r})]$ at different times during the measurement sequence, corresponding to
    $\tau_e=\tau_d=960\,\mu\text{s}$, and a sample displacement of $\delta z = 1$~\AA.
    (b) The data were acquired using the sequence shown in (a), where $2\times$SME4 sequences were used in the encoding portion, with each SME4 designed to generate a Rabi gradient evolution time of $\Delta\tau=480\,\mu\text{s}$. From the fit to the data, we determine the ratio $\delta y/\delta z = 6.5$.
    For reference, two other curves have been included that show the variation in signal amplitude for different ratios of $\delta y / \delta z$. The calculations clearly indicate that for lateral displacements of order 1~nm, a higher ratio $\delta y/\delta z$ primarily increases the decay rate with respect to $V_P$ of the modulation envelope of the interference signal, without significantly affecting the modulation frequency; the modulation frequency, on the other hand, is primarily determined by $\delta z$. The asymmetry in the response along the $y$ and $z$ directions is caused by the relative magnitude of the Rabi field gradients near the center of the CFFGS, which is at least a factor of 100 larger in the $z$ direction than in the $y$ and $x$ directions.
    (c) In-phase and quadrature data were acquired for $V_P=0\,V$ and $V_P=7.2\,V$ with a fixed encoding time of $\tau_e=480\,\mu\text{s}$ for different decoding times. A single SME4 with $\Delta\tau=480\,\mu\text{s}$ was used for the encoding and decoding parts of the sequence.
    The displacement results in a shift in the phase of both signal quadratures.
    From the fit to the data, we determine displacement corresponding to $V_P=7.2\,V$ to be $\delta z = 0.81\pm0.07$~\AA\, and $\delta y = 9.72\pm 1.34$~\AA. The value of $\delta y$ determined from the fit is consistent with the value $\delta y = 1.0\,\text{nm}$ determined by optical interferometry (see Sec.~VII of Ref.~\cite{Supplements}.)}
    \label{fig:Fig6}
\end{figure*}

In Sec.~\ref{NMRd_18X_grating}, we performed phase-sensitive NMRd measurements of a nanometer-scale periodic modulation of the $z$-axis magnetization with sub-\AA ngstrom precision. In this section, we describe an interferometric detection protocol that enables us to measure a real-space displacement of the InP sample in the $z$ direction with a precision of 0.07~\AA. The protocol, shown schematically in Fig.~\ref{fig:Fig6}(a), utilizes the symmetric magic echo (SME) \cite{Boutis2003} NMR sequence to decouple the P-P and P-In interactions, thereby extending the coherence time of the $^{31}$P spins up to 12.8~ms. In Sec.~IV(D) of Ref.~\cite{Supplements}, we describe a modification to the SME4 that allows us to evolve the spins under the Rabi field gradient for a variable amount of time $\Delta\tau$ for phase encoding. By extending the spin coherence time into the millisecond range and by utilizing the large Rabi field gradients of order $2\times 10^5$~T/m, we encode a helical phase winding in the $xz$ plane with an average wavelength as short as a few \AA ngstroms, allowing us to detect displacements of the InP sample with picometer precision.

The protocol starts by encoding a helical winding for a time $\tau_e$.
The density matrix for a spin at position $\mathbf{r}$ after encoding is $\rho_1=U\left[u(\mathbf{r}),\tau_e\right]\rho_0U^{\dagger}\left[u(\mathbf{r}),\tau_e\right]$, where $U\left[u(\mathbf{r}),\tau\right]=\exp\left[{-i2\pi\tau\sigma_y u(\mathbf{r})/2}\right]$. During this time, a constant voltage is applied to the PT [inset in Fig.~\ref{fig:Fig6}(b)], which translates the InP sample with respect to the CFFGS, thereby slightly shifting the local field experienced by the $^{31}$P spins during encoding.
The PT is retracted to its equilibrium position by zeroing the voltage.
During retraction, $2\times$SME4 sequences with $\Delta\tau=0$ are applied, i.e., no gradient evolution, which refocus the homonuclear dipolar and $\sigma_z$ evolution during the time that the PT returns to its equilibrium position. The duration of the $2\times$SME4 sequences is 1.8~ms, which is chosen to be substantially longer than the $\sim 300\,\mu\text{s}$ mechanical response time of the PT.
In the decoding phase, the inverse unitary  $U^{\dagger}\left[u(\mathbf{r}+\delta\mathbf{r}),\tau_d \right]=\exp\left[i2\pi\tau_d u(\mathbf{r}+\delta\mathbf{r})\sigma_y/2\right]$ is applied for a time $\tau_d$ at the new location of the InP sample.
The density matrix at the end of the sequence is $\rho_2=\delta U \rho_0\delta U^\dagger \propto \sigma_z \cos \varphi(\mathbf{r})  +\sigma_x \sin \varphi(\mathbf{r})$, with $\delta U = \exp[-i\varphi(\mathbf{r})\sigma_y/2]$. The differential phase $\varphi(\mathbf{r})=2\pi\left[u(\mathbf{r})\tau_e-u(\mathbf{r}+\delta\mathbf{r})\tau_d\right]$ results from the interference of the encoding and decoding modulations separated by $\delta\mathbf{r}$.
The measured signal quadratures at the end of the sequence are
\begin{equation}
    \begin{bmatrix}
        s_I(\tau_e,\tau_d,\delta\mathbf{r})\\
        s_Q(\tau_e,\tau_d,\delta\mathbf{r})
    \end{bmatrix} \propto \int d^3r\,\rho(\mathbf{r})
    \begin{bmatrix}
        \cos\varphi(\mathbf{r})\\
        \sin\varphi(\mathbf{r})
    \end{bmatrix}
    \label{eq:DisplacementDiffraction}
\end{equation}
Here, $\rho(\mathbf{r})$ is the effective spin density at location $\mathbf{r}$ (see Appendix~\ref{Appendix}).

Figure~\ref{fig:Fig6}(b) shows a plot of the in-phase data acquired as a function of the voltage $V_P$ applied to the PT, using the sequence shown in Fig.~\ref{fig:Fig6}(a), with $\tau_d=\tau_e = 960~\mu\text{s}$. The modulation wavelength varies from $2.1$~\AA\, to $5.8$~\AA\, within the measured volume of the sample. 
To extract the $z$-displacement from the data, we conduct a least-squares fit \cite{sivia2006} using Eq.~\eqref{eq:DisplacementDiffraction}, where we assume that $\delta \mathbf{r}\propto V_P$ and $\delta \mathbf{r} = \delta y~\hat{\mathbf{y}}+\delta z~\hat{\mathbf{z}}$, with the fit parameters being the PT coefficient $\delta z/V_P$ and $\delta y/\delta z$.
The displacement in the $y$ direction is included because the PT used for the measurements was poled in the $y$ direction; hence $\delta y(V_P) \gg \delta z(V_P)$. The effective spin density $\rho(\mathbf{r})$ was calculated using our model for the sample geometry (Sec.~III of Ref.~\cite{Supplements}).
For details regarding the characterization of the PT, see Sec.~VII of Ref.~\cite{Supplements}.
The resulting fit is indicated by the solid line in Fig.~\ref{fig:Fig6}(b), corresponding to a displacement $\delta z/V_P = 0.15\pm 0.01$~\AA/V. The PT calibration was used to derive the top horizontal axis of Fig.~\ref{fig:Fig6}(b).

Finally, we conduct a set of measurements in which we keep the encoding time and the voltage step applied to the PT constant, and vary the decoding time. Fig.~\ref{fig:Fig6}(c) shows the in-phase and quadrature data acquired for $V_P=0$~V and $V_P=7.2$~V with $\tau_e=480\,\mu\text{s}$.
To extract the sample displacement, we fit to the data using Eq.~\eqref{eq:DisplacementDiffraction}, with the fit parameters being $\delta z$ and $\delta y$ [solid lines in Fig.~\ref{fig:Fig6}(c)].
The fit yields a sample displacement of $\delta z = 0.81\pm 0.07$~\AA.
In Sec.~VIII of Ref.~\cite{Supplements}, we also provide an alternative fitting method, which utilizes the measured $p(u)$ distribution instead of the geometric model of the sample. Consistent with the previous method, the fit yields $\delta z = 0.85\pm 0.07$~\AA.

Using the PT displacement calibration found from Fig.~\ref{fig:Fig6}(b), we would expect the displacement corresponding to $V_P=7.2$~V to be $\delta z=1.04\pm 0.04$~\AA. Although the calculated displacements from the two measurements are in reasonable agreement, the 20\% difference in $\delta z$ could be caused by systematic errors in the different methods used for determining the displacement corresponding to $V_P=7.2$~V. In particular, to determine the PT displacement for the data in Fig.~\ref{fig:Fig6}(b), we needed to assume a particular functional form for the piezo characteristic $\delta\mathbf{r}(V_P)$. For the data in Fig.~\ref{fig:Fig6}(c), however, we find the displacement directly for one particular value of $V_P$, without any assumptions on $\delta\mathbf{r}(V_P)$. It is therefore possible that a small nonlinear component in $\delta\mathbf{r}(V_P)$ could be responsible for the observed difference.

\begin{figure}[!htbp]
    \centering
    \includegraphics[width = \columnwidth]{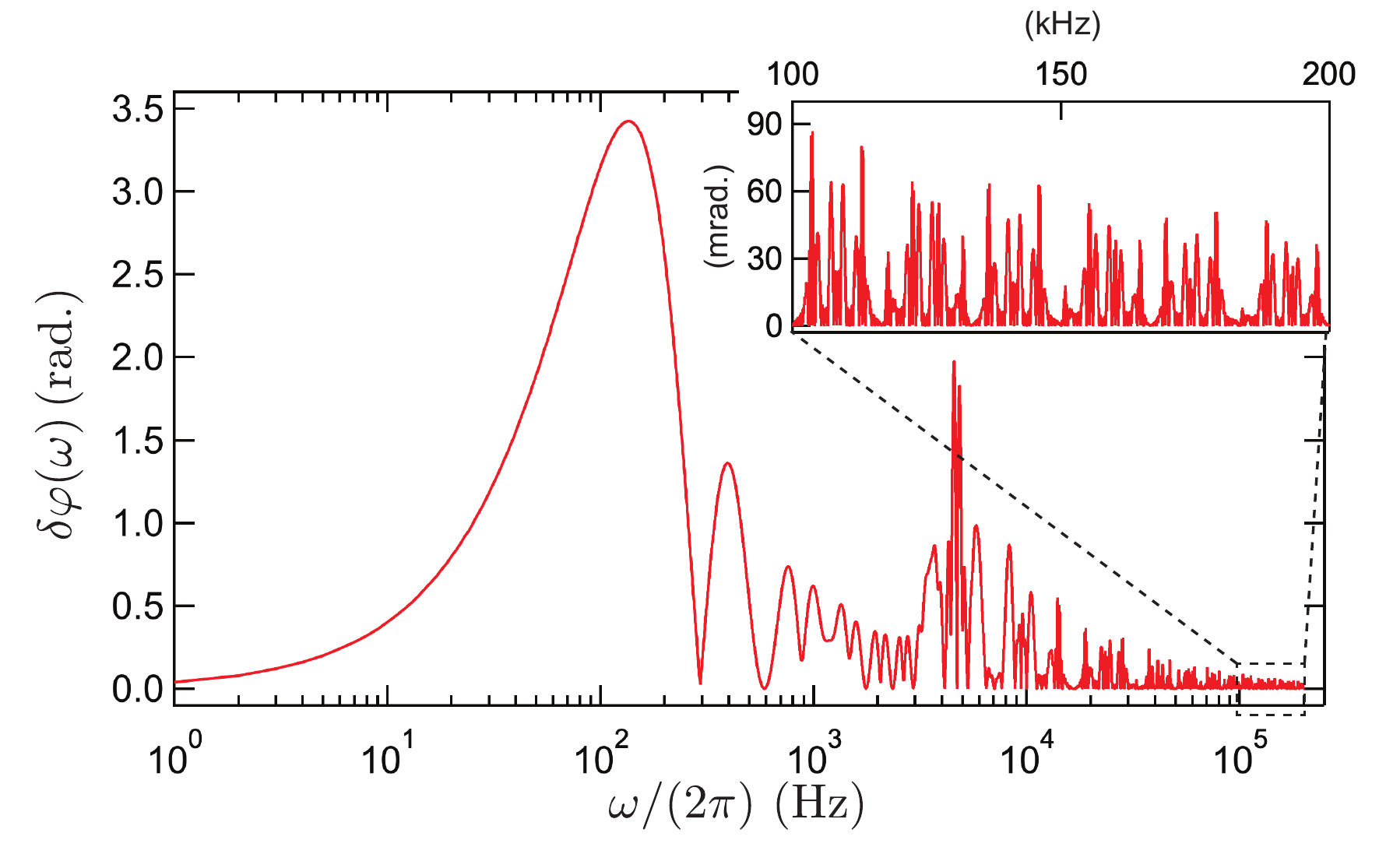}
    \caption{Phase error $\delta\varphi(\omega)$ calculated by integrating the phase accumulated by a $^{31}$P spin oscillating at frequency $\omega$ with a peak amplitude of 1~\AA\, in the $z$ direction and experiencing the sample-averaged Rabi-field gradient of ${\partial B_1/\partial z} = 1.8$~G/nm. For the calculation, we use the sequence in Fig.~\ref{fig:Fig6}(a) with $\tau_{\text{ME}}=212\,\mu\text{s}$ and $T=5.1$~ms.}
    \label{fig:Fig.7}
\end{figure}

For measurements that require large encoding wavevectors, such as those that would be needed for crystallographic NMRd, the mechanical stability of the sample becomes an important consideration. To characterize the response to mechanical motion, we calculate the root-mean-square integrated phase error $\delta\varphi(\omega)$, caused by the motion of the sample relative to the gradient source during the encoding sequence.
As an example, Fig.~\ref{fig:Fig.7} shows a plot of $\delta\varphi(\omega)$ calculated for the sequence shown in Fig.~\ref{fig:Fig6}(a). We can clearly see that at low frequencies, even \AA ngstrom-scale motion can introduce large phase errors. In particular, the main peak near 100~Hz is caused by a 1~\AA-peak motion near the frequency $1/(2T)$, where $T$ is the duration of the entire sequence; the smaller peak near 5~kHz represents the motion at $1/\tau_{\text{ME}}$, where $\tau_{\text{ME}}$ is the duration of a single magic echo sequence. Overall, however the phase error decreases substantially for frequencies $\omega/(2\pi)>100\,\text{kHz}$, near the SiNW mechanical resonance, where the \AA ngstrom-scale motion of the oscillator caused by thermal fluctuations would be of concern.
Phase errors can be further minimized by applying feedback cooling to reduce the motion of the oscillator \cite{Poggio2007}. For the measurements reported here, the encoding direction is perpendicular to the oscillation direction of the SiNW, which greatly reduces phase errors caused by the resonant motion of the oscillator.
To ensure that such errors are negligible, we compare the measured the signal after applying the sequence in Fig.~\ref{fig:Fig6}(a) at zero piezo voltage, with $\tau_e = \tau_d = 0$ and $\tau_e = \tau_d= 960 \ \mu$s -- the longest encoding time used in this work.
The result shows a small $11\pm 10\%$ reduction at $960 \ \mu$s encoding, which indicates sufficient mechanical stability in the reported experiments.

\section{Conclusions}
%In this work, we presented two experiments that demonstrate phase-sensitive NMRd measurements of $^{31}$P spins with sub-\AA ngstrom precision.
In this work, we presented two experiments that utilize the large encoding wavevectors generated in a nanoMRI setting to realize phase-sensitive position measurement of $^{31}$P spins with sub-\AA ngstrom precision. These results demonstrate new capabilities for studying material structure and quantum phenomena by extending the the spectroscopic and imaging capabilities of NMR to the atomic scale. Although the concept of NMRd was first envisioned for the study of material structure, it can be applied more generally to study the dynamics of spatially-periodic spin correlations. For example, experiment presented in Sec.~\ref{NMR_Interferometer} can be readily adapted to study quantum transport by replacing the storage period with an evolution under an effective Hamiltonian. In fact, such experiments have been performed to study dipolar spin diffusion of $^{19}$F spins in $\text{CaF}_2$ on the micron scale \cite{Zhang1998}. Here, we have demonstrated the ability to generate encoding wavevectors that are nearly a factor of $10^4$ times larger than these previous works, which could be used to probe spin transport on length scales as short as the lattice spacing, where quantum phenomena could become important. In addition, the interferometric method introduced in this work to measure displacement could be extended to study molecular motion on the \AA ngstrom scale using Fourier imaging techniques. More generally, atomic-scale NMRd could be applied to Hamiltonian engineering applications by providing local control of spins in periodic spin systems \cite{Ajoy2013}, and for probing correlations in nuclear spin chains, where novel quantum many-body correlations have been observed \cite{Wei2018}.

The application of  NMRd to study three-dimensional crystal structure will require devices capable of generating highly uniform field gradients. 
Upcoming experiments focusing on NMRd crystallography will utilize a new design that combines the CFFGS with four additional current carrying paths designed to generate highly uniform three-dimensional field gradients of order $3\times 10^4$~T/m with a maximum variation of 0.5\% in a $\sim(100~\text{nm})^3$ volume, suitable for crystallographic NMRd measurements. Using these gradients, $^1$H spins separated by 3~\AA, for example, would form a DE at an encoding time of 2.6~ms, which is readily achievable using dynamical-decoupling NMR sequences, such as the SME. This capability could be used to study the structure of organic nanocrystalline materials, such as protein nanocrystals, that are of great interest in structural biology.

\section{Acknowledgements}
This work was undertaken thanks in part to funding from the U.S. Army Research Office through Grant No.
W911NF1610199, the Canada First Research Excellence
Fund (CFREF), and the Natural Sciences and Engineering Research Council of Canada (NSERC). The University of Waterloo's QNFCF facility was used for this work. This infrastructure would not be possible without the significant contributions of CFREF-TQT, CFI, Industry Canada, the Ontario Ministry of Research and Innovation and Mike and Ophelia Lazaridis. Their support is gratefully acknowledged. H.H. would like to thank  F. Flicker, S. H. Simon and H. S. Røising for insightful and inspiring discussions. R.B thanks D. G. Cory and A. Cooper for useful discussions.

H. Haas and S. Tabatabaei contributed equally to this work.

%%% References %%%
% \nocite{*}
% \bibliography{bibliography}
\providecommand{\noopsort}[1]{}\providecommand{\singleletter}[1]{#1}%

\appendix

\section{Spin Detection Protocol}\label{Appendix}
\begin{figure}[h!]
    \centering
    \includegraphics[width = \columnwidth]{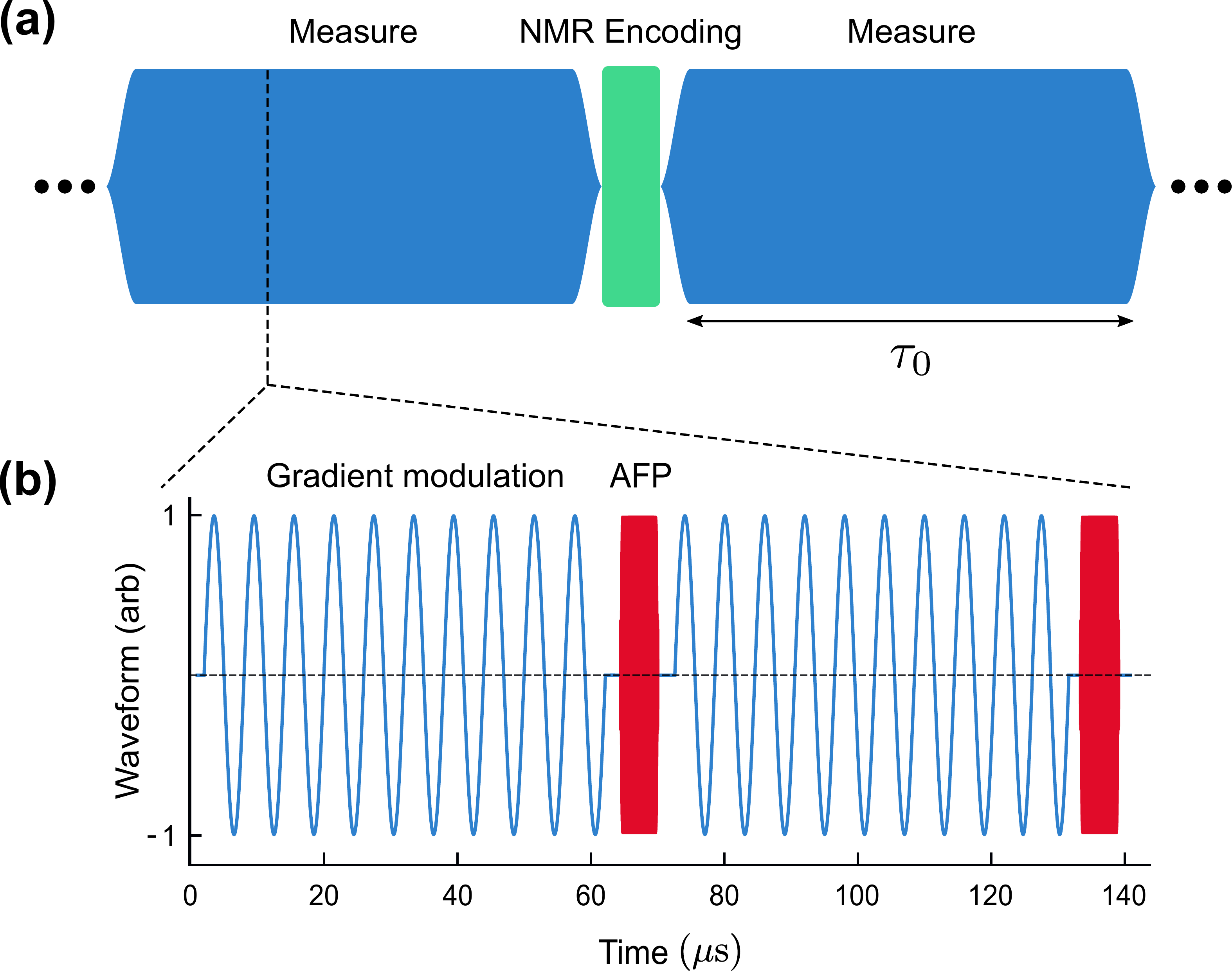}
    \caption{MAGGIC spin detection protocol. (a) General overview of the protocol. The blue regions indicate the spin readout period during which the MAGGIC waveform is applied. The correlation signal [Eq.~\eqref{eq:MAGGICsignal}] is constructed from the average correlation of two consecutive force measurements made before and after the NMR encoding sequence.
    (b) The MAGGIC waveform primitive.}
    \label{fig:Fig8}
\end{figure}

Here, we briefly discuss our measurement protocol, while referring to \cite{Rose2018} and the supplements therein for further details. 
We use the MAGGIC protocol \cite{Nichol2012}, shown in Fig.~\ref{fig:Fig8}, to detect the force generated by the $z$-axis magnetization from the spin ensemble.
The protocol is comprised of successive applications of a measurement block, separated by an NMR encoding block containing the spin control operations of interest -- e.g., the sequences in Fig. \ref{fig:Fig4}(c) and Fig. \ref{fig:Fig6}(a) for our two experiments.
The duration of each measurement block $(\tau_0=1.0~\text{s})$ is chosen to be shorter than the correlation time of $\tau_m=2.9$~s of the statistical fluctuations for the $^{31}$P spins, measured using the MAGGIC protocol.

The MAGGIC waveform [Fig.~\ref{fig:Fig8}(b)] applied in each measurement block consists of two periods, during which the gradient $\partial B_z/\partial y$ is modulated at the resonant frequency of the SiNW.
To avoid spurious electrical couplings to the SiNW, we ensure that the modulation waveform has no Fourier component at $\omega_c$ by applying a periodic $\pi$ phase shift to the gradient modulation. An adiabatic full passage (AFP) pulse is applied synchronously with the phase shifts, which creates a force from the spins that is on resonance with the SiNW.
The average force correlation between two consecutive measurement blocks is
\begin{align}
    s \propto \int d^3r~n(\mathbf{r}) G_R^2(\mathbf{r}) \chi\big(u(\mathbf{r})\big) \langle \sigma_z \rangle_{\mathbf{r}},\label{eq:MAGGICsignal}
\end{align}
where $n(\mathbf{r})$ is the spin density, $G_R(\mathbf{r})=\partial B_z(\mathbf{r})/\partial y$ is the peak amplitude of the readout gradient, $\langle \sigma_z \rangle_{\mathbf{r}}$ is the expectation value of $\sigma_z$ at positon $\mathbf{r}$ at the end of the NMR encoding block, and $\chi(u)$ is a filter function quantifying the performance of the AFPs at Rabi frequency $u$ \cite{Rose2018}. 
From Eq.~(\ref{eq:MAGGICsignal}), we identify the effective spin density as $\rho(\mathbf{r}) = n(\mathbf{r}) G_R^2(\mathbf{r}) \chi\big(u(\mathbf{r})\big)$.

In the case where the NMR encoding block sequence is only selective in Rabi frequency [e.g., the sequence in Fig.~\ref{fig:Fig4}(c)], $\langle \sigma_z\rangle_{\mathbf{r}}$ is the same for spins experiencing equal Rabi frequencies, which simplifies Eq.~\eqref{eq:MAGGICsignal} to $s \propto \int du~p(u) \langle \sigma_z \rangle_u$, where $p(u)$ is the effective Rabi-frequency distribution.
Given the approximate one-to-one relationship between $z$ and $u$ in our sample volume, $p(u)$ can be written as
\begin{align}
    p(u) \approx \frac{\partial z(u)}{\partial u} \int_{-\infty}^\infty dx \int_{-\infty}^\infty dy~ \rho\big(x,y, z(u)\big).
\end{align}

For the measurement blocks, we utilized the 2.3 Rabi-cycle AFP presented in Ref.~\cite{Tabatabaei2021}.
For the experiment in Sec.~\ref{NMRd_18X_grating}, the AFP was 4.8~$\mu$s long, with a near-unity fidelity over $480 \text{ kHz} \leq u \leq 960 \text{ kHz}$, and for the experiment in Sec.~\ref{NMR_Interferometer}, the pulse was rescaled to be 5.7~$\mu$s to target the Rabi range $400 \text{ kHz} \leq u \leq 800 \text{ kHz}$.
In the latter experiment, we also restrict the detection volume to $u\leq 650$~kHz by including an additional waveform in the beginning of the NMR encoding block.
Acting as a low-pass filter, the waveform evolves spins aligned with the $z$-axis to the transverse plane for $u>650$ kHz, which then dephase due to transverse relaxation.
Spins experiencing $u\leq650$~kHz are not affected
(see Ref.~\cite{Supplements}).

%%% Supplements %%%
\clearpage
\clearpage


\section*{Supplementary material}

# Supplementary Information for Nanometer-Scale Nuclear Magnetic Resonance Diffraction with Sub-Angstrom Precision

Holger Haas[¶,1,2,*] Sahand Tabatabaei[¶,1,2] William Rose,[3] Pardis Sahafi,[1,2]
Michèle Piscitelli,[1,2,†] Andrew Jordan,[1,2] Pritam Priyadarsi,[1,2] Namanish
Singh,[1,2] Ben Yager,[1,2,‡] Philip J. Poole,[4] Dan Dalacu,[4] and Raffi Budakian[1,2,§]

[1]*Department of Physics and Astronomy, University of Waterloo, Waterloo, ON, Canada, N2L3G1*
[2]*Institute for Quantum Computing, University of Waterloo, Waterloo, ON, Canada, N2L3G1*
[3]*Department of Physics, University of Illinois at Urbana-Champaign, Urbana, Illinois 61801, USA*
[4]*National Research Council of Canada, Ottawa, Ontario, Canada, K1A 0R6*

## I. SAMPLE ATTACHMENT

The attachment of an indium phosphide nanowire (InPNW) sample to the silicon nanowire (SiNW) force detector relies solely on the attractive van der Waals force between the nanowires. To attach the sample, we used a micromanipulator with a sharp glass pipette tip under an optical microscope for breaking an InPNW from its growth substrate and bringing it into contact with the SiNW, allowing the two nanowires to adhere. No adhesives were used in any part of the attachment process. The InPNW was attached $\sim 3-4$ $\mu m$ away from the tip of the SiNW with a slight overhang. Fig. S1 shows the two nanowires before and after the attachment, as well as a scanning electron microscope image of the InPNW array.

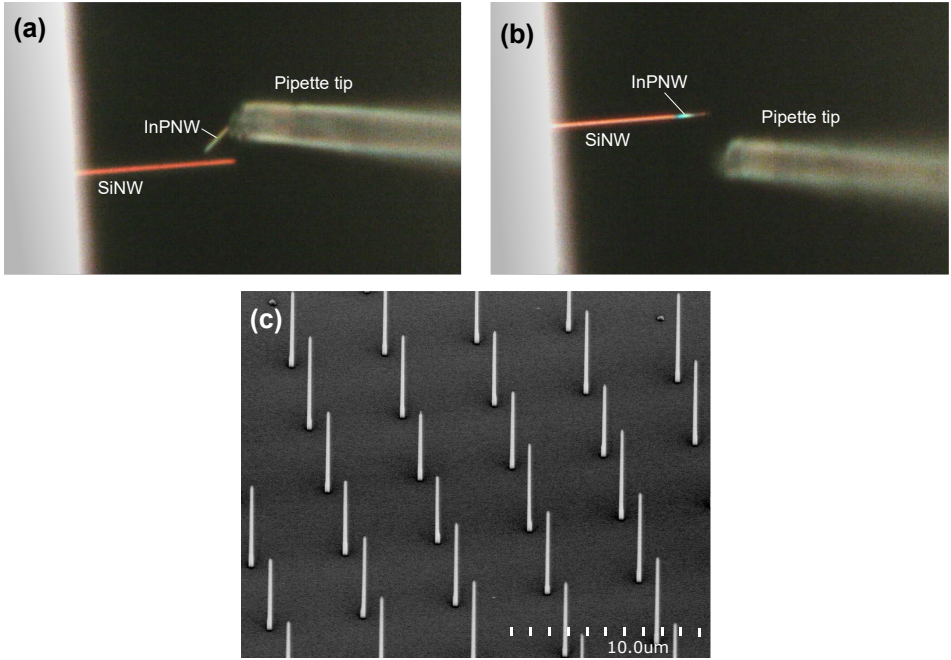

FIG. S1. The indium phosphide nanowire (InPNW) sample (a) before and (b) after the attachment using the micromanipulator. (c) Scanning electron microscope image of the InPNW array.

---

\* Present address: IBM Quantum, IBM T.J. Watson Research Center, Yorktown Heights, NY 10598, USA
† Present address: Clarendon Laboratory, Department of Physics, University of Oxford, OX1 3PU, UK
‡ Present address: Oxford Instruments NanoScience, Abingdon, Oxfordshire, OX13 5QX, UK
§ rbudakian@uwaterloo.ca
¶ H. Haas and S. Tabatabaei contributed equally to this work.

## II. CFFGS FIELD DISTRIBUTION

We used radio frequency (RF) currents of 70-mA peak amplitude driven through the CFFGS (Fig. S2) to generate spatially-varying RF magnetic fields. The field profiles are calculated using COMSOL finite element simulations under the quasi static approximation since the dimensions of the CFFGS are negligible compared to the wavelength of the utilized currents. The plots in Fig. S3 show contours of constant Rabi frequency $u(\mathbf{r}) = \gamma\sqrt{B_x^2(\mathbf{r}) + B_y^2(\mathbf{r})}/(4\pi)$, Rabi field gradient $\partial B_1(\mathbf{r})/\partial z$ and readout gradient $G_R(\mathbf{r}) = \partial B_z(\mathbf{r})/\partial y$. The plot also includes the $z$-dependencies of the above-mentioned quantities along the axis passing through the center of the constriction ($x = y = 0$).

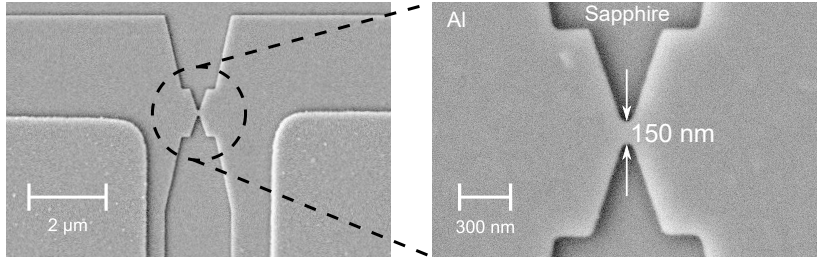

FIG. S2. Scanning electron microscope image of the CFFGS device.

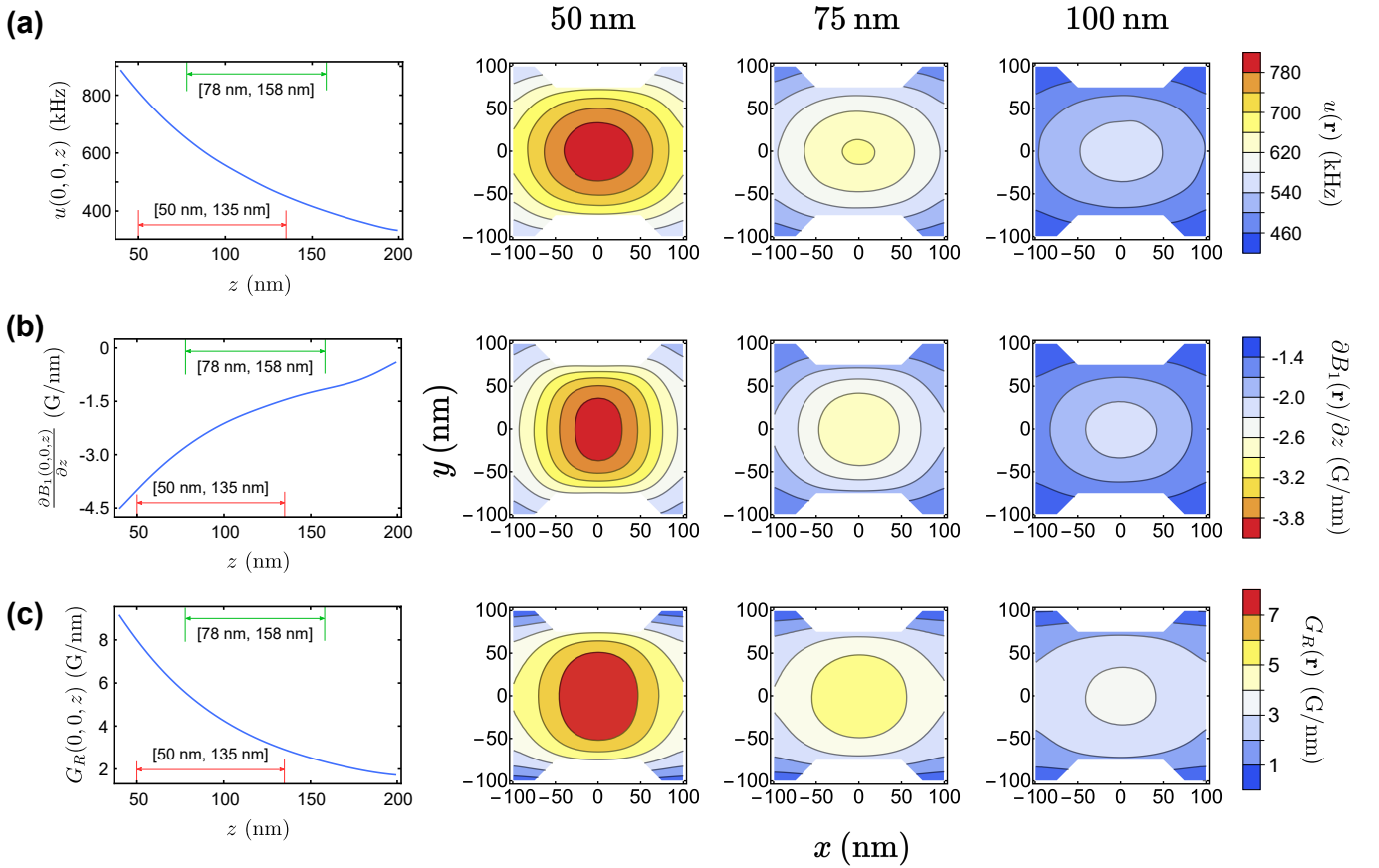

FIG. S3. Simulated field profile in the vicinity of the current-focusing field gradient source (CFFGS) corresponding to a 70 mA peak amplitude currents driven through the CFFGS. Each row indicates the contours of constant (a) Rabi frequency, (b) Rabi $z$-gradient and (c) readout gradient. The first column indicates the $z$-dependence of said quantities along the $x = y = 0$ line, and the other columns correspond to contour plots 50 nm, 75 nm and 100 nm above the center of the CFFGS. The red and green arrows in the first column correspond to the detection region for the nanoscale NMRd (Sec. IV) and NMR interferometry experiment (Sec. V) discussed in the main text, respectively.

## III. SAMPLE GEOMETRY CHARACTERIZATION

To characterize the sample geometry, we perform 1D imaging measurements by phase encoding using the Larmor and Rabi field gradients generated by the CFFGS. We then construct a model of the InP sample, and calculate the time-domain signal using the simulated field gradient profile discussed in Sec. II. In the optimization process, we vary the 2D profile of the nanowire in the $yz$ plane, and generate a 3D geometry by assuming rotational symmetry about the nanowire axis. In addition, a tilt angle of the nanowire axis with respect to the $z$ axis in the $yz$ plane is also included to account for any misalignment. The parameters of the 2D geometry and the tilt axes are varied to achieve the best correspondence between the measured and calculated signals.

The time-domain signals are measured by applying either static or resonant pulses to encode using Larmor-frequency or Rabi-frequency gradients, respectively. The measurement protocol and the associated parameters are shown in Fig. S4(b,c). All measurements are made by varying the pulse amplitude, while keeping the pulse width fixed. Bandwidth-limited shaped pulses used for Larmor-frequency encoding are applied during the free-evolution period of the SME4 sequence (Sec. IV D), as shown in Fig. S4(c). The $\pi/2$ rotation pulses are taken to be the optimized pulse discussed in Sec. IV C, with its duration rescaled to 15.5 $\mu$s to target the [350 kHz, 740 kHz] Rabi frequency range. For the Rabi and Larmor encoding measurements, we define the effective encoding time as: (Rabi) $\tau_{\mathrm{eff}} = \tau_1 I_{\mathrm{pk}}/I_0$, (Larmor) $\tau_{\mathrm{eff}} = 8b\tau_0 I_{\mathrm{pk}}/I_0$, where $b = 0.824$ is a numerical factor that scales area of the pulse to the area of a square pulse with the same amplitude, and $I_0 = 70$ mA. $\tau_{\mathrm{eff}}$ represents the width of a single encoding pulse applied with a peak amplitude of $I_0$. We report encoding times in terms of $\tau_{\mathrm{eff}}$ to have a common reference for all measurements.

The 1D imaging measurements were made by encoding $^{31}$P and $^1$H spins. We attribute the presence of $^1$H spins to a $\sim$2-nm thick layer of protons on the surface of the InPNW. Although we do not know the origin of this layer, based on the short $T_2 = 12 \mu$s, it is likely a dense network of protons, possibly from water. By measuring the frequency distribution of both spin species, we have two independent measurements to corroborate the tip model. Fig. S5 shows the measured Rabi and Larmor encoding data for $^{31}$P and $^1$H spins, as well as the calculated signals. The calculations are done using the geometry that best fits the measured 1D data [Fig. S4(a)]. We find that the tilt axis of the InPNW that best fits the data is nearly zero (the InPNW axis is aligned parallel to the $z$ direction).

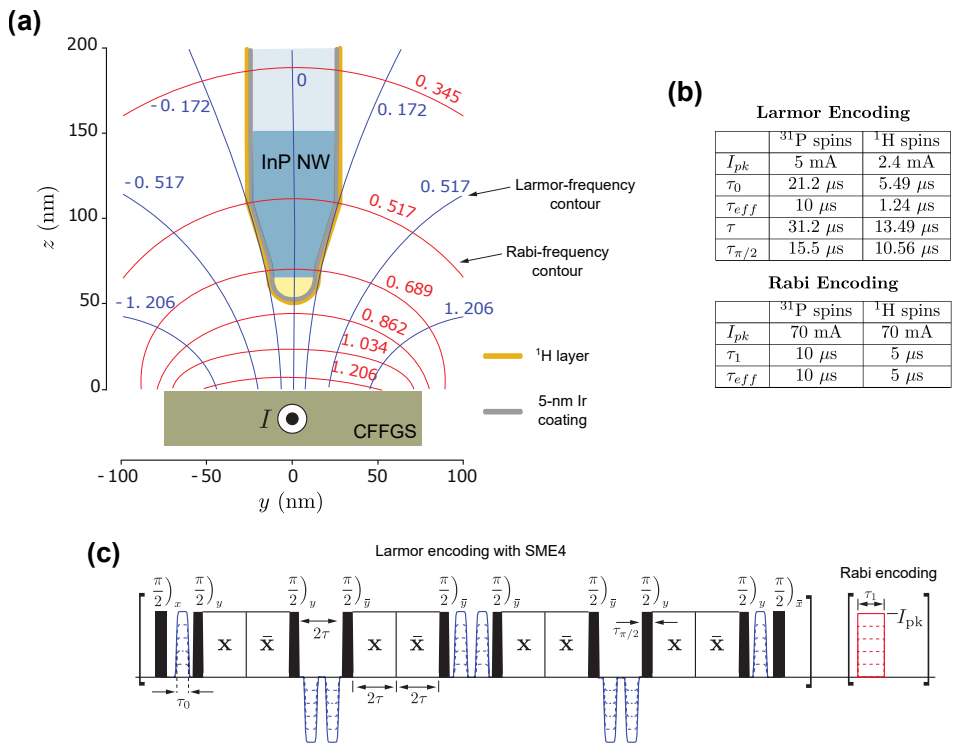

FIG. S4. (a) Contours of constant resonance offset and Rabi frequency referenced to a peak current amplitude of 70 mA, in units of MHz. The figure shows the optimized sample geometry and a cross section through the center of the 150-nm wide constriction. (b) The parameters used for the 1D Larmor and Rabi imaging measurements. (c) For the 1D imaging measurements that utilized resonance offset (Larmor) gradients, pulses were applied during the free-evolution times in the SME4 sequence. For the 1D imaging measurements that utilized the Rabi frequency gradients, a single resonant pulse of length $\tau_1$ was applied.

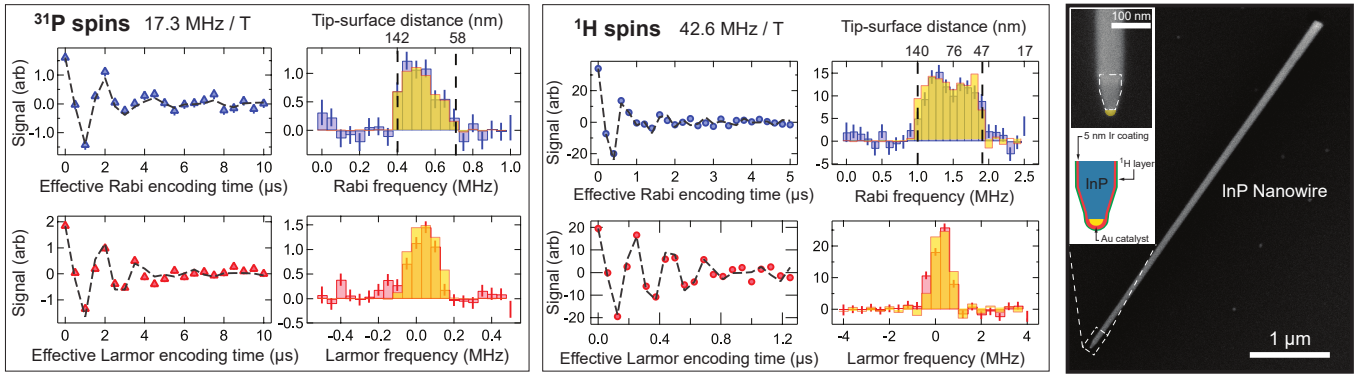

FIG. S5. (Left) The measured time-domain Larmor and Rabi encoding data vs. effective encoding time for $^{31}$P and $^{1}$H spins are shown with the data points. The dashed lines indicate the calculated signal for the tip geometry shown in the inset on the right. The discrete cosine transform of the data set is shown as the solid colored histograms. The frequency distributions corresponding to the calculated geometry is shown as the yellow overlay. The two vertical dashed lines indicate the upper and lower Rabi frequencies used for the measurements. (Right) SEM of a representative InPNW. The calculated region for the nanowire used for the measurements is shown as with the dashed lines on an SEM of a representative InPNW. Note, this is not the nanowire used for the measurements, however we did SEM a number of nanowires, all of which had a similar tapered region. The 20-nm diameter Au catalyst particle used in the InPNW growth is indicated on the SEM in false color.

## IV. SPIN CONTROL SEQUENCES

In this section, we describe the details of the spin control sequences referenced in the main text.

### A. Adiabatic Half-Passages

The adiabatic half-passages (AHPs) used in our work were designed using the numerical optimization protocol presented in [1]. We parametrize the instantaneous Rabi frequency and resonance offset waveforms as $u_{\max} \tanh[a_x(t)]$ and $\Delta\nu_{\max} \tanh[a_z(t)]$, respectively, where

$$
\begin{cases}
a_x(t) = \sum_{n=1}^{20} x_n \left[ 1 - \left(1 - \dfrac{t}{T}\right)^{2n} \right] \\
a_z(t) = \sum_{n=21}^{40} x_n \left(1 - \dfrac{t}{T}\right)^{2n-41}
\end{cases}
\tag{S1}
$$

Here, $\{x_n\}$ are the optimization parameters, $T$ is the pulse duration, and $u_{\max}$, $\Delta\nu_{\max}$ are the maximum Rabi frequency and resonance offset, respectively. The initial and final states in the optimization were chosen to be $|\uparrow\rangle$ and $(|\uparrow\rangle + |\downarrow\rangle)/\sqrt{2}$, respectively, with the optimization including metrics for maximizing the overlap of the final and target states, as well as the overall adiabaticity of the transfer. For the details of the optimization protocol and the associated metrics, see Sec. II of [1]. The optimized pulse is 12-$\mu$s long, works over a Rabi range of $u_{\max} \in [300 \text{ kHz}, 860 \text{ kHz}]$, and has a maximum resonance offset of $\Delta\nu_{\max} = 3.1$ MHz.

### B. Band Inversion Pulses

It is well known [2, 3] that the effective magnetic field in the FM-frame [1], where the $z$ component of the field amounts to instantaneous resonance offset, can be used as a spin-locking field for adiabatic control operations such as the AFPs and AHPs used in this work. It is known that by engineering the spin to follow the effective field trajectory, the operation can be made robust to a wide variety of perturbation Hamiltonians, such as resonance offsets and RF field inhomogeneities.

It is, however, possible to engineer spin-locking fields that are not robust to a specific perturbation. This is exploited here for engineering Rabi frequency selective control sequences. Consider the evolution under the Hamiltonian

$H(t; f) = 2\pi u \, b_x(t; f)\sigma_x/2$, where $b_x$ is a periodic function defined over one cycle as

$$b_x(t; f) = 1 + 2ft^2\left(ft^2 - 1\right), \qquad -\frac{1}{\sqrt{2f}} \le t \le \frac{1}{\sqrt{2f}}, \tag{S2}$$

and extended to other times via $b_x(t; f) = b_x\left(t + n\sqrt{2/f}; f\right)$ for $n \in \mathbb{Z}$. Here, $f > 0$ is a free parameter. The Hamiltonian generates the propagator $U(t) = \exp\left[-i\int_0^t dt' H(t')\right]$, which is equal to identity for Rabi frequency $u = n\sqrt{2f}15/23$, after one $b_x$ cycle ($t = \sqrt{2/f}$). At the same time, the 0th order average Hamiltonian theory (AHT) [4] integral of $\sigma_z$ over the same cycle

$$\int_0^{\sqrt{2/f}} dt_1 \, U^\dagger(t_1)\sigma_z U(t_1), \tag{S3}$$

is, in general, non-zero. Therefore, a spin at $u = n\sqrt{2f}15/23$ starting from a $\sigma_x$ eigenstate, i.e., aligned with the spin-lock field, will rotate under any non-zero $\sigma_z$ perturbation to $H(t)$. Conversely, the spins that do not experience $u = n\sqrt{2f}15/23$ and start from a $\sigma_x$ eigenstate will remain in that state in the presence of small $\sigma_z$ perturbations, as a non-identity $U\left(t = \sqrt{2/f}\right)$ ensures that the effect of the perturbation is averaged out.

The observations in the previous paragraph suggest an obvious protocol for generating Rabi frequency selective rotations. By first applying an AHP on the spins, such that they end up in a $\sigma_x$ eigenstate, and thereafter applying a Hamiltonian $H(t; f) = 2\pi u \, b_x(t; f)\sigma_x/2 + 2\pi\epsilon\sigma_z/2$, where $\epsilon$ is a small resonance offset, a Rabi frequency selective rotation can be engineered the rate of which is proportional to the value of $\epsilon$. Furthermore, the amplitude of $\epsilon$ also determines the range of Rabi frequencies that the selective rotation affects – larger $\epsilon$ results a larger 'slice' in $u$ being rotated. Finally, by introducing a chirp to $b_x(t; f)$, a continuous range of different Rabi frequencies can be targeted. The spins can be realigned with the $z$ axis by following the $b_x$ waveform with another AHP.

In this work, we used control sequences that applied the principles described above in 3 separate instances: for inverting the 100 kHz-wide $u$ band in Fig. 3 of the main text, for encoding the 18-band diffraction grating demonstrated in Fig. 4 of the main text and for implementing the low-pass filter when collecting the data in Fig. 6 of the main text. All three operations start and end with AHPs to ensure the correct spin orientation. We give the exact waveform parameters for each of these experiments below. In all cases we generated a Hamiltonian

$$H(t; f) = a(t)\frac{\sigma_x}{2} + \Delta\omega(t)\frac{\sigma_z}{2} = 2\pi u b_x(c_2 t^2 + c_1 t; f)\frac{\sigma_x}{2} + \Delta\omega(t)\frac{\sigma_z}{2}, \tag{S4}$$

with $\Delta\omega(t)$ always shaped with a hyperbolic tangent function to smoothly introduce the resonance offsets. We also define the response of the Rabi-selective operation as the expectation value $g(u) = \langle\sigma_z\rangle_u$ at the end of the sequence (including the AHPs) for a spin starting from $|\uparrow\rangle$. The $c_1, c_2$ and $f$ parameters, as well as the exact shape of $\Delta\omega(t)$, were adjusted to yield the desired action on the spins in simulation. The chirp rate $c_2$ and the maximum value of $\Delta\omega(t)$ both determine the 'sharpness' of the $g(u)$ profile. In practice, we need to keep $\Delta\omega(t)$ sufficiently large such that the inhomogeneous lineshape ($\propto 1/T_2^*$) of the spins would not significantly affect the desired action.

### 1. 100 kHz Band Inversion

The 100 kHz band inversion waveforms are

$$a_{\text{band}}(t) = 2\pi u \, b_x\left(t \times 3.511, 778 + t^2 \times 43.354 \text{ Hz}; \, 0.029, 777 \text{ MHz}^2\right), \tag{S5}$$

$$\frac{\Delta\omega_{\text{band}}(t)}{2\pi} = \frac{16 \text{ kHz}}{4}\left[1 + \tanh\left(\frac{t - 241.243 \ \mu s}{80 \ \mu s}\right)\right]\left[1 - \tanh\left(\frac{t - 7,450.684 \ \mu s}{80 \ \mu s}\right)\right], \tag{S6}$$

over a pulse duration $0 \le t \le 7,692.928 \ \mu s$. The $b_x(t; f)$ waveform is defined by Eq. (S2). An illustration of the waveforms, as well as the instantaneous target Rabi frequency $u_t$ for the chirped $b_x(t; f)$ profile are shown in Fig. S6(a-c). The resulting $g(u)$ profile can be seen in Fig. 3(c) of the main text. When simulating the $g(u)$ profile we account for the inhomogeneous lineshape in exactly the same way as in calculating the signal of Eq. (S15).

### 2. Diffraction Grating

The diffraction grating magnetization profile $g(u)$, as shown in Fig. 4(a) of the main text, is generated by sweeping $u_t$ for the chirped $b_x(t; f)$ waveform from 440 to 805 kHz, while periodically turning on non-zero resonance offset over

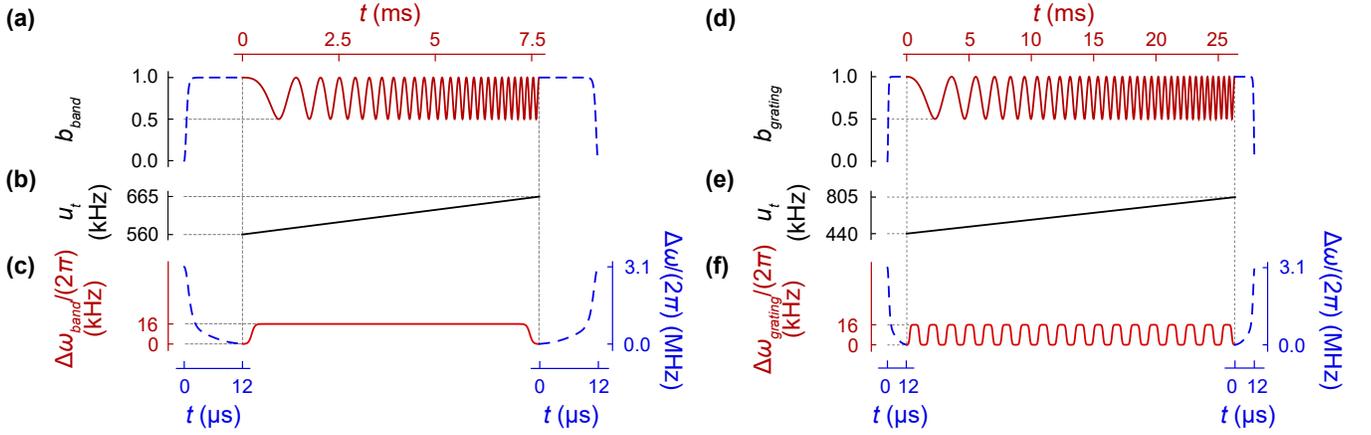

FIG. S6. (a) Illustration of the $a(t)$ waveform in Eq. (S4) that was used for generating the 100 kHz-wide band inversion. The sequence starts and ends with the 12 $\mu$s-long AHPs of Sec. IV A (blue dashed lines), whereas its middle part is the 7.692 ms-long function (red continuous line) given by Eq. (S5). The oscillations of $b_{band}$ are illustrative rather than the actual function. (b) Rabi frequency $u_t$ targeted by the instantaneous oscillation period of Eq. (S5). (c) Resonance offset function $\Delta\omega(t)$ in Eq. (S6) for the 100 kHz band inversion. The AHP (blue dashed lines) sweeps onto resonance from 3.1 MHz above the spin Larmor frequency (axis on the right), whereas the maximum resonance offset of Eq. is 16 kHz (red continuous line, axis on the left). (d) Sketch of $a(t)$ in Eq. (S4) used for generating the 18-band diffraction grating. The 26.311 ms-long waveform function given by Eq. (S7) (red continuous line) is sandwiched between two AHPs (blue dashed lines). The oscillations of $b_{band}$ are illustrative. (e) $u_t(t)$ for Eq. (S7). (f) $\Delta\omega(t)$ in Eq. (S8) for the diffraction grating sequence. The 18 sweeps with hyperbolic tangent shapes to $\Delta\omega/(2\pi) = 16$ kHz values generate the 18 regions with $g(u) \approx -1$ in Fig. 4(a) of the main text.

the pulse duration $0 \leq t \leq 26,311.460$ $\mu$s. We illustrate the waveforms used, including the AHPs at the start and end, in Fig. S6(d-f). The exact waveforms used were

$$a_{\text{grating}}(t) = 2\pi u \ b_x(t \times 2.769, 909 + t^2 \times 43.354 \text{ Hz}; \ 0.029, 777 \text{ MHz}^2), \tag{S7}$$

$$\frac{\Delta\omega_{\text{grating}}(t)}{2\pi} = \sum_{k=1}^{18} \frac{16 \text{ kHz}}{4} \left[ 1 + \tanh\left( \frac{t + 1,235.698 \ \mu s - k \times 1,474.922 \ \mu s}{80 \ \mu s} \right) \right] \tag{S8}$$

$$\times \left[ 1 - \tanh\left( \frac{t + 478.237 \ \mu s - k \times 1,474.922 \ \mu s}{80 \ \mu s} \right) \right].$$

Again, when simulating $g(u)$ for Fig. 4(a) of the main text, we account for the inhomogeneous lineshape.

### 3. 650 kHz Low-Pass Filter

For the experiments reported in Sec. V of the main text, we engineered a 'low-pass' filter for setting the $z$-magnetization of spins with Rabi frequency ($u$) greater than 650 kHz to 0. This was achieved by applying a waveform

$$a_{\text{filter}}(t) = 2\pi u \ b_x \left( t \times 3.945,810 + t^2 \times 370.920 \text{ Hz}; \ 0.029,777 \text{ MHz}^2 \right), \tag{S9}$$

$$\frac{\Delta\omega_{\text{filter}}(t)}{2\pi} = \frac{16 \text{ kHz}}{4} \left[ 1 + \tanh\left( \frac{t - 250 \ \mu s}{83.333 \ \mu s} \right) \right] \left[ 1 - \tanh\left( \frac{t - 2,250.690 \ \mu s}{83.333 \ \mu s} \right) \right] \tag{S10}$$

over $0 \leq t \leq 2,500.690$ $\mu$s, which was again sandwiched between two AHPs. The resulting $g(u)$ profile is shown in Fig. S7. It can be seen that unlike the previous $g(u)$ profiles that take values nominally at around -1 and 1, in this case we set $g(u) \approx 0$ for 650 kHz $\leq u \leq$ 900 kHz. This ensures that the signal in the measurement is only collected from spins with $u \leq 650$ kHz, which is further enforced by applying the filter multiple times.

## C. $\pi/2$ Rotation Pulse

Given the non-uniform Rabi frequency distribution generated by the CFFGS, it is non-trivial to generate global unitary operations for all spins in the measurement volume. We numerically engineer global unitaries over a broad

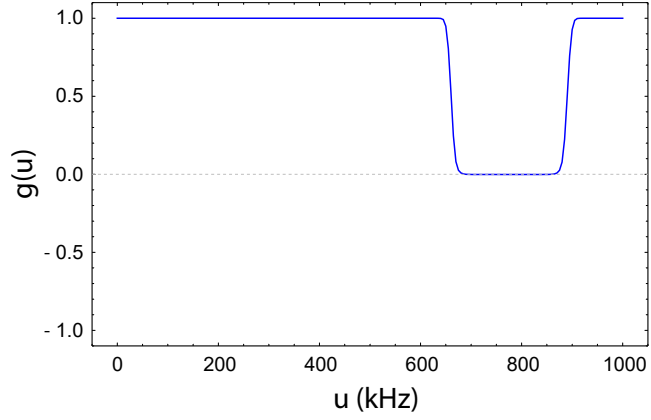

FIG. S7. $g(u) = \langle \sigma_z \rangle_u$ after applying a single low-pass filter function determined by waveform in Eq. (S9) to a spin in the $|\uparrow\rangle$ state experiencing a particular Rabi frequency $u$.

range of Rabi frequencies experienced by our sample spins using the gradient-based optimization protocol presented in detail in [5]. The protocol has been previously used in [6] for nanoMRI experiments. In addition to optimizing the final unitary, the protocol allows for the inclusion of various robustness criteria against perturbation Hamiltonians, all for a broad range of Rabi frequencies. It also enables us to restrict the bandwith of the control waveform and account for its distortions due to experimental transfer function (TF) of the control electronics and transmission lines.

For the $\pi/2$ pulse used in our experiments, we optimized a $T = 16.2$ $\mu$s long pulse. We denote the in-phase and quadrature envelope functions of the pulse by $a_x(t)$ and $a_y(t)$, respectively. These waveforms were parametrized with Legendre polynomials of order 34, which are passed through a Gaussian filter to constrain the waveform bandwidth. The optimization accounts for the TF, which was measured prior to the experiments [Fig. S8(b,c)]. We utilize this numerically optimized pulse for all global $\pi/2$ rotations used in this work. The rotating-frame Hamiltonian for a spin at Rabi-frequency $u$ is given by $H(t) = 2\pi u[\tilde{a}_x(t)\sigma_x + \tilde{a}_y(t)\sigma_y]/2$, where $\tilde{a}_x(t)$ and $\tilde{a}_y(t)$ are the distorted waveforms at the output of the experimental TF, assuming a carrier frequency of 52 MHz. The corresponding propagator is given by $U(t) = \text{Texp}[-i \int_0^t dt' H(t')]$, with Texp denoting the time-ordered exponential. The protocol is set up to minimize the infidelity of the final unitary with respect to $\exp(-i\pi\sigma_x/4)$:

$$\varphi_0 \equiv 1 - \frac{1}{4}\left|\text{Tr}[U(T)e^{i\pi\sigma_x/4}]\right|^2,$$ (S11)

along with a robustness metric for resonance offsets (e.g. heteronuclear couplings to In spins)

$$\varphi_z \equiv \frac{\|\mathcal{D}_z\|^2}{\max(\|\mathcal{D}_z\|^2)} = \frac{1}{2T^2}\text{Tr}[\mathcal{D}_z^\dagger \mathcal{D}_z],$$ (S12)

where $\mathcal{D}_z \equiv U(T)\int_0^T dt\ U^\dagger(t)\sigma_z U(t)$ is the associated first-order perturbation integral, and a robustness metric for homonuclear dipolar interactions

$$\varphi_{\text{dip}} \equiv \frac{\|\mathcal{D}_{\text{dip}}\|^2}{\max(\|\mathcal{D}_{\text{dip}}\|^2)} = \frac{1}{24T^2}\text{Tr}[\mathcal{D}_{\text{dip}}^\dagger \mathcal{D}_{\text{dip}}],$$ (S13)

where, similarly,

$$\mathcal{D}_{\text{dip}} \equiv [U(T)\otimes U(T)]\int_0^T dt\ [U^\dagger(t)\otimes U^\dagger(t)]\big(2\sigma_z\otimes\sigma_z - \sigma_x\otimes\sigma_x - \sigma_y\otimes\sigma_y\big)[U(t)\otimes U(t)],$$ (S14)

is the corresponding perturbation integral. All three metrics were minimized over the Rabi-frequency range 355 kHz $\leq u \leq 709$ kHz. Figure S8 shows the optimized waveform envelopes $a_x(t)$ and $a_y(t)$, along with the calculated single-spin final unitary, dipolar and $\sigma_z$ robustness metrics. The figure also includes the metrics for the distorted waveforms $\tilde{a}_x(t)$ and $\tilde{a}_y(t)$ at the output of the TF.

We experimentally validated the pulse by measuring the ensemble-average $z$-magnetization after repeated back-to-back applications [see Fig. S8(g)]. The measurement was conducted with the sample Rabi-frequency distribution of

Fig. 6(a) in the main text. Since the signal decays to $1/e$ after $\sim$120 pulses, we estimate that the operation works with $\sim$99% accuracy over our Rabi range. Figure S8(g) also includes simulations of the $\pi/2$ pulse on an ensemble of non-interacting spins, accounting for the measured Rabi-frequency distribution $p(u)$ as well as the Lorentzian inhomogenous lineshape $p_{\mathrm{inh}}(\nu) \propto [1 + (2\pi T_2^* \nu)^2]^{-1}$, with $T_2^* = 70\ \mu s$. Denoting the final unitary of the $\pi/2$ pulse at Rabi frequency $u$ and resonance offset $\nu$ by $\mathcal{U}(u, \nu)$, the simulated signal is

$$s_n \propto \sum_u \sum_\nu p(u) p_{\mathrm{inh}}(\nu) \mathrm{Tr}\Big[\sigma_z\, \mathcal{U}^n(u, \nu)\sigma_z[\mathcal{U}^n(u, \nu)]^\dagger\Big], \tag{S15}$$

where the sum over $\nu$ samples the $[-20\ \mathrm{kHz}, 20\ \mathrm{kHz}]$ range at 21 points, and the sum over $u$ takes 16 samples from the $[330\ \mathrm{kHz}, 760\ \mathrm{kHz}]$ range.

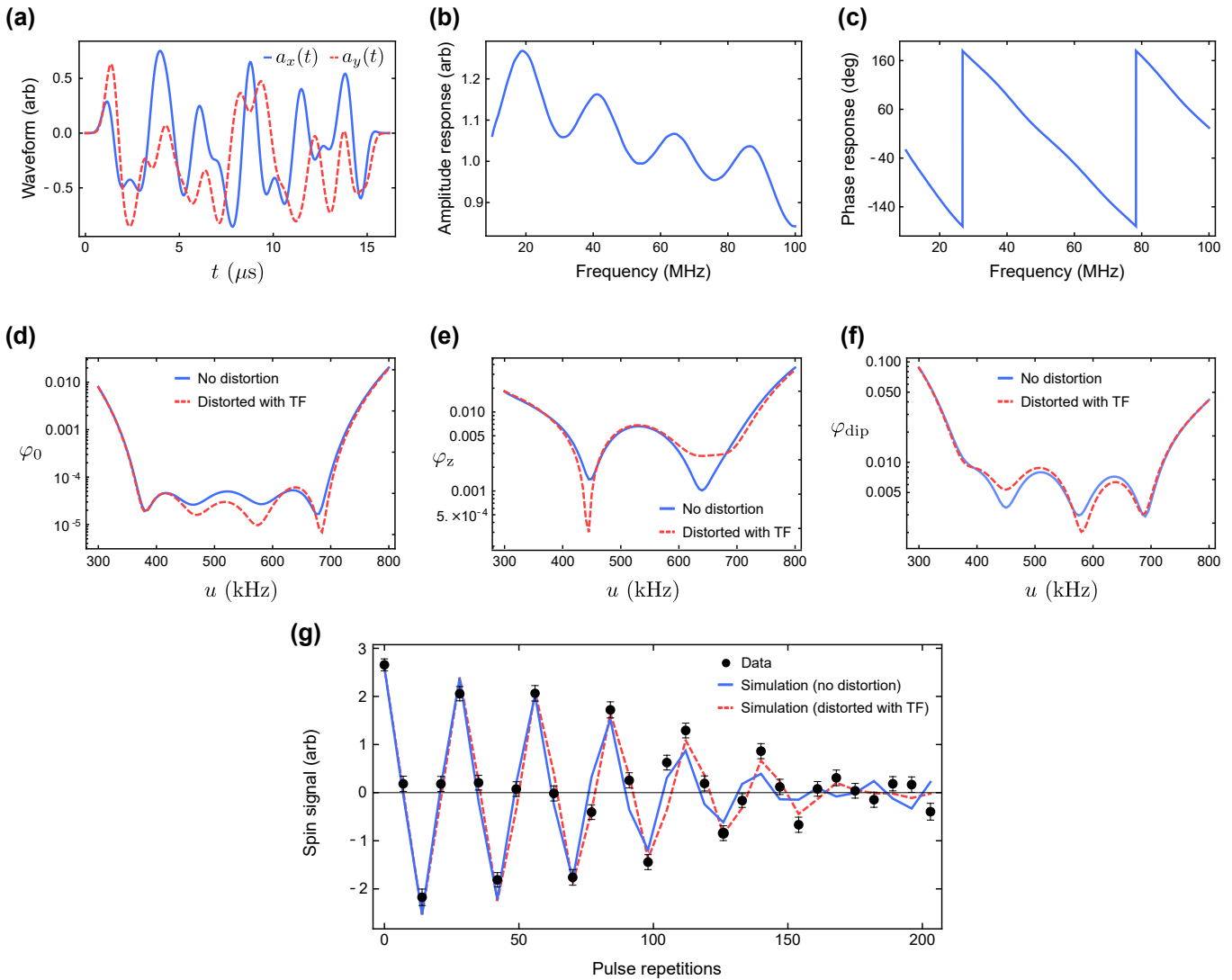

FIG. S8. (a) In phase and quadrature envelope functions of the numerically optimized $\pi/2$ pulse. (b, c) Amplitude and phase of the measured transfer function (TF) for the electronics in our setup. (d) Final unitary, (e) $\sigma_z$ robustness and (f) dipolar robustness metrics for the optimized pulse with and without the TF distortion. (g) Experimental data after repeated back-to-back applications of the optimized $\pi/2$ pulse, along with simulations of the original and distorted waveforms.

### D. Magic Echo Sequences

In this section, we present the details of the symmetric magic echo (SME) sequence used in the NMR interferometry experiment. The basic magic echo (ME) primitive [outlined block in Fig. S9(a)] is comprised of two free evolution

blocks of length $\tau$, two $\pi/2)_y$ rotations (taken to be the optimized pulse of Sec. IV C), and a rotary echo (RE) of length $4\tau$ in the middle. The ME sequence is known to refocus homonuclear dipolar couplings as well as all $\sigma_z$-type Hamiltonians, e.g. , resonance offsets, chemical shifts and heteronuclear interactions. In our experiments, we utilize the SME4 sequence [see Fig. S9(a)]– a composite sequence of ME blocks robust to pulse imperfections [7]. The advantage of dynamical decoupling using ME-based sequences is that they allow for simultaneous evolution under both static and RF field gradients for Fourier encoding, making them ideal for imaging applications. Figure S9(b) depicts a modification to the SME4 sequence with a time offset in the RE to encode phase under the Rabi-field gradient. For a spin at Rabi frequency $u$, an SME4 sequence asymmetrized by $\Delta\tau/8$ in each RE results in a final unitary $U_{\text{SME4}}(\Delta\tau) = \exp(-i2\pi u\Delta\tau\sigma_z/2)$ in the rotating frame. We use SME4 sequences with $\tau = 31~\mu s$, meaning that for a single SME4 sequence, $\Delta\tau \leq 496~\mu s$.

A complication we observed when encoding Ångström-scale wavelengths with SME sequences is that small transients from the RF circuitry can cause errors in the encoded phase, which can in turn introduce a systematic error to the measured displacement in the NMR interferometry experiment. For each RE section of the SME4, a transient causing a fractional change in amplitude $\epsilon(t)$ generates a phase error $\int_0^{4\tau} dt\, \epsilon(t) \times u = \tilde{\tau} u$ for a spin at Rabi frequency $u$, where $\tilde{\tau}$ is an effective error in the encoding time. Therein, in an encode-decode sequence [e.g., Fig. 6(a) of the main text], the effect of transients can be modeled as an effective difference in encoding and decoding times $\delta\tau = \tilde{\tau}_e - \tilde{\tau}_d$. To demonstrate the scale of this phase error, consider encoding and then decoding phase on a spin at $u = 500$ kHz for $\tau_e = \tau_d = 480~\mu s$ – the smallest encoding time used in this work. In this case, the spin is ideally driven for 240 full revolutions around the Bloch sphere and back, meaning that only a $\delta\tau/\tau_e = 0.2\%$ error can result in a large overall overrotation of $180°$.

To diagnose this issue, we conducted an experiment in which we encode for a fixed time $\tau_e$ and measure the spins after decoding for a variable time $\tau_d$. Since the signal $s(\tau_d)$ has the form $s(\tau_d) \propto \int du\, p(u) \cos[2\pi u(\tau_e - \tau_d + \delta\tau)]$, the overrotations break the even symmetry of $s(\tau_d)$ about $\tau_d = \tau_e$ in this measurement. The red squares in Fig. S10(a, b) depict two such measurements with $\tau_e = 480~\mu s$ and $\tau_e = 960~\mu s$, indicating an asymmetric signal and hence the presence of overrotations. We minimize the RF transients by filling the free evolution times in the SME4 using an off-resonant drive whose amplitude is matched to the rest of the sequence, with a frequency $\Delta\nu = 40$ MHz above resonance [blue regions in Fig. S9(b)]. This 'filler drive' equalizes the power throughput of the RF circuitry in time, thereby reducing any thermal transients. An important side effect of the filler drive is the introduction of an effective resonance offset due to AC Stark shift [8]. In the reference frame rotating at the Larmor frequency, the Hamiltonian of a filler drive at Rabi frequency $u$ and resonance offset $\Delta\nu$ is $H_f(t) = 2\pi u[\cos(2\pi\Delta\nu t)\sigma_x + \sin(2\pi\Delta\nu t)\sigma_y]/2$, which in average Hamiltonian theory (AHT) [4] can be equated with an effective Hamiltonian $\bar{H}_f \approx -\pi u^2\sigma_z/(2\Delta\nu)$ in the free evolution times of the SME4 sequence. In our case, the effective resonance offset is averaged out by the SME4 sequence. Repeating the diagnostic measurement for SME4 sequences with the filler show a symmetric signal and hence negligible phase errors for both encoding times [Fig. S10(a, b)].

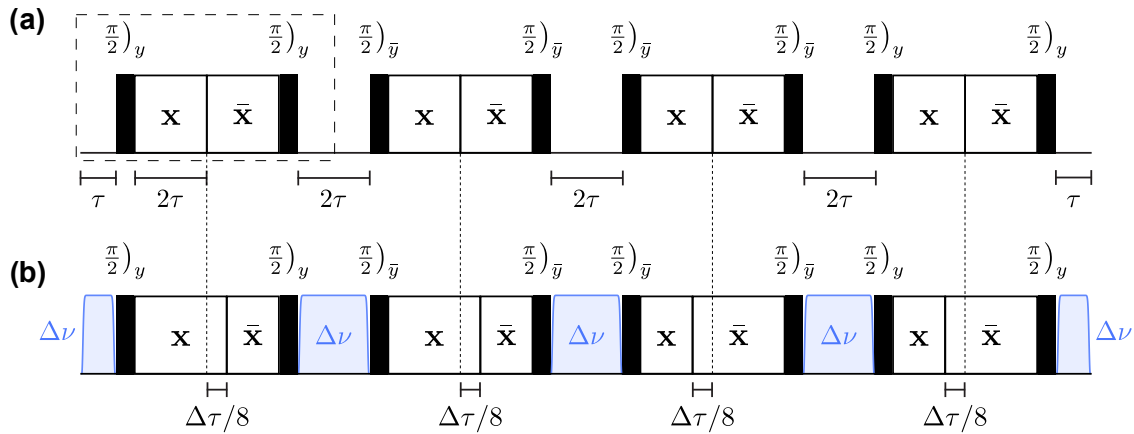

FIG. S9. (a) The SME4 sequence for dynamical decoupling. The outlined block represents a single magic echo primitive. (b) Modified SME4 sequence for phase encoding with Rabi-field gradients. The blue regions indicate the off-resonant filler drive for suppressing transients from the RF electronics. In our experiments, $\Delta\nu = 40$ MHz.

To demonstrate dynamical decoupling, we measure spins after repeated applications of the SME4 sequence with $\Delta\tau = 0$ [Fig. S10(c)]. From an exponential fit to the data, we measure an extended spin coherence time of $12.8\pm1.0$ ms and $15.9 \pm 0.5$ ms with and without the filler drive, respectively.

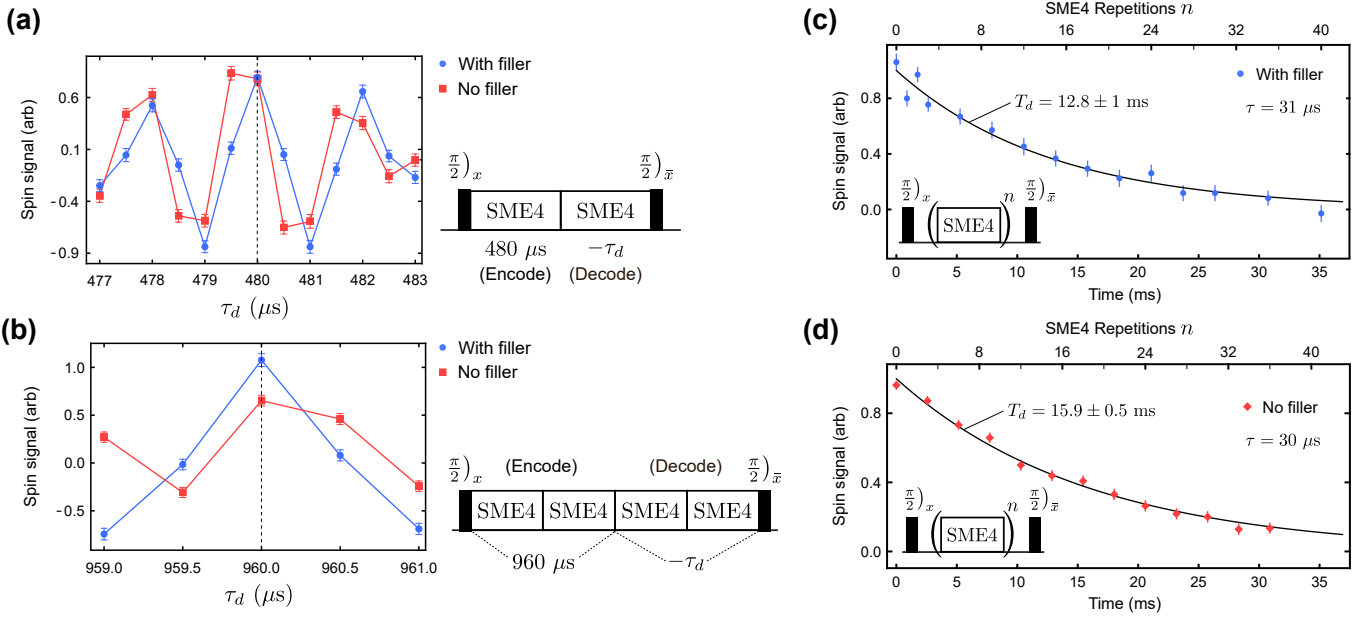

FIG. S10. (a, b) Diagnostic data with and without the off-resonant filler drive for determining overrotations coming from pulse transients, along with the associated pulse sequences. Introducing the off-resonant filler drive makes the signal amplitude symmetric around $\tau_d = 480$ µs in (a) and $\tau_d = 960$ µs in (b). (c, d) Back-to-back applications of symmetric ($\Delta\tau = 0$) SME4 sequences for line narrowing. The SME4 with and without the filler were applied with $\tau = 31$ µs and $\tau = 30$ µs, respectively.

## V. $^{31}$P SPIN CHARACTERIZATION

We characterized the $^{31}$P spin relaxation times under both continuous drive and free evolution with four separate measurements (Fig. S11). We measured $T_2^*$ with a Ramsey experiment [Fig. S11(a)] in which we tipped the spins into the $xy$ plane using an AHP, freely evolved them for time $\tau$, applied an inverse AHP, and measured the longitudinal magnetization as a function of $\tau$. From the fit to the data, we determined $T_2^* = 70 \pm 7$ µs.

To obtain $T_2$, we used a $\sigma_z$-decoupling sequence comprised of periodic spin inversions which refocus any $z$-rotations such as the heteronuclear couplings to the In spins. The spin flips were separated by $\delta\tau = 20$ µs. The decay was measured by tipping the spins into the transverse plane with an AHP, followed by $n$ inversions, applying the inverse AHP, and measuring the spin signal as a function of the total free evolution time $\tau = n\delta\tau$. From the data shown in Fig. S11(b), we obtain a Gaussian fit with $T_2 = 271 \pm 14$ µs.

We also measured the spin dephasing time under continuous resonant drive, ($T_{2\rho}$) with a rotary echo (RE) experiment. Starting from the $z$ axis, the spins precess around the $x$ axis for time $\tau/2$ under a constant drive, and are then reverted with a drive of the opposite phase for the same duration. Measuring the longitudinal magnetization as a function of $\tau$ gives a decay curve describing the dephasing under the continuous drive. We find that the dephasing best matches a Gaussian curve with time constant $T_{2\rho} = 568 \pm 13$ µs [Fig. S11(c)]. Since the homonuclear dipolar Hamiltonian is effectively rescaled by $-1/2$ under the RE, we expect $T_{2\rho} \approx 2T_2$, which is consistent with our measured values for $T_2$ and $T_{2\rho}$. Because all Rabi encoding pulses in our experiments are done with encoding times much less than $T_{2\rho}$, we can safely neglect this relaxation mechanism in this work.

To obtain the $T_{1\rho}$ decay time, we applied a spin-locking sequence in which the spins were flipped onto the $x$ axis using an AHP and then spin-locked for time $\tau$ with a resonant $\mathbf{x}$ drive. After taking the spins back to the $z$ axis by applying the inverse AHP, we measured the longitudinal magnetization for different values of $\tau$. From the fit to the data [Fig. S11(d)], we find $T_{1\rho} = 4.0 \pm 0.4$s. Finally, our coarse characterization of the $T_1$ decay established that it was greater than 5 s.

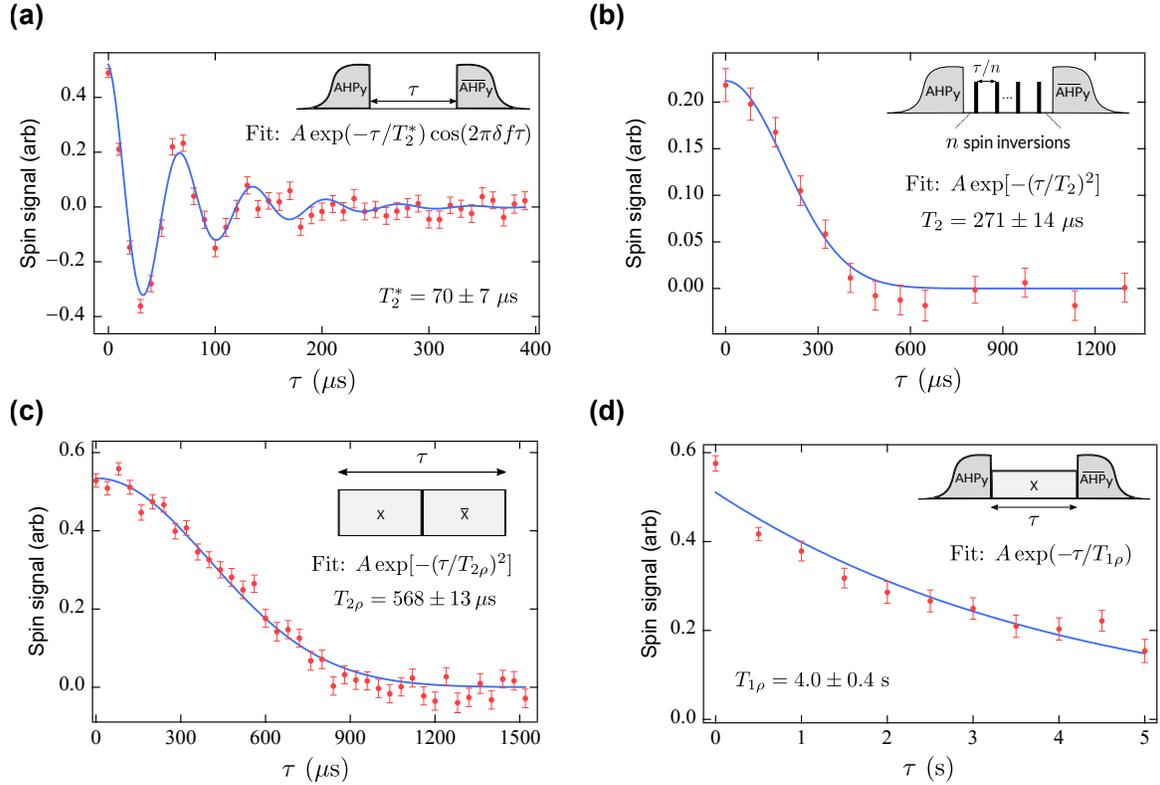

FIG. S11. Relaxation time measurements including (a) $T_2^*$ (Ramsey), (b) $T_2$ ($\sigma_z$ decoupling), (c) $T_{2\rho}$ (rotary echo) and (d) $T_{1\rho}$ (spin-lock). The insets show the pulse sequence used for each experiment.

## VI. ESTIMATORS AND UNCERTAINTIES FOR 18×DIFFRACTION GRATING EXPERIMENT

In this section, we detail our exact methodology for estimating the period of the grating ($\Omega$) in its 2 positions as well as the $u$-space translation ($\Delta u$) between these positions for the 18-band diffraction grating experiment. We work exclusively with in-phase and quadrature measurements of the first diffraction echo for grating positions 1 and 2, the corresponding data sets are shown in Fig. 4 of the main text. The data presented results from calculating the average force correlation of ∼2500 to ∼3000 measurement blocks in the MAGGIC spin detection protocol.

In the following subsections, we determine estimates for the period $\Omega$ and translation $\Delta u$ as well as their uncertainties using the measured data of Fig. 4(a) in the main text. While doing this analytically, we attempt to apply statistical estimator theory [9] reasonably rigorously. Nevertheless, we will generally do it implicitly so as to avoid introducing unnecessary jargon. We also verify the analytically-calculated uncertainties in $\Omega$ and $\Delta u$ through Monte Carlo sampling. It will be made explicitly clear that our analysis is entirely agnostic to the way the nuclear spin magnetization grating is created, and exploits no information related to it. The only major assumption in our analysis is that the variations in the effective spin density $p(u)$ are significantly slower than the period of the grating. Using the CFFGS field simulations of Sec. II, we tie the $u$-space quantities and their uncertainties to real space values.

Because each diffraction echo measurement in Fig. 4(d, f) of the main text is an average of a large number of MAGGIC force correlations, we assume the measurements at different times $t \in \mathcal{T}$, with $\mathcal{T} = \{39\ \mu s, 39.5\ \mu s, ..., 58.5\ \mu s, 59\ \mu s\}$ to be realizations of independent normally-distributed random variables $\{S_t\}$ and $\{Q_t\}$ for the in-phase and quadrature measurements, respectively. The normally-distributed nature of $\{S_t\}$ and $\{Q_t\}$ arises due to the central limit theorem, despite the distribution of the individual correlation measurements being non-Gaussian. We have verified this assertion by analyzing the statistics of individual correlation measurements as well as the statistics of correlation measurement averages over subsets of all measurements.

In the following, when necessary, we distinguish between the two grating positions with a Roman numeral upper index; e.g., $S_t^{\mathrm{II}}$ denotes the random variable corresponding to the measurement of the in-phase NMRd signal at position 2 for time $t$. The expectation values of $S_t$ and $Q_t$ are denoted as $\mathrm{E}\,(S_t)$ and $\mathrm{E}\,(Q_t)$, whereas their variances $\mathrm{E}\,(S_t^2) - \mathrm{E}\,(S_t)^2$ and $\mathrm{E}\,(Q_t^2) - \mathrm{E}\,(Q_t)^2$ are denoted as $\mathrm{Var}\,(S_t)$ and $\mathrm{Var}\,(Q_t)$, respectively. We denote the observed measurement outcomes – i.e., the data values shown in Fig. 4(d, f) in the main text – by $\{s_t\}$ and $\{q_t\}$; these values

serve as our best estimates for the expectation values $\{E(S_t)\}$ and $\{E(Q_t)\}$ and will be used as such when necessary throughout this section. Finally, because $s_t$ and $q_t$ are calculated by averaging a random sample of individual force correlation measurements, we can estimate $\mathrm{Var}(S_t)$ and $\mathrm{Var}(Q_t)$ as the variance of the sample mean of the correlations, i.e. the variance of individual correlations divided by the number of measurements. We denote these estimates of $\mathrm{Var}(S_t)$ and $\mathrm{Var}(Q_t)$ as $(\sigma_t)^2$ and $(\varphi_t)^2$, respectively. The error bars in Fig. 4(d, f) of the main text represent the various $\sigma_t$ and $\varphi_t$ values.

### A. Estimating the Grating Period

Here, we estimate the period $\Omega$ of the spin magnetization grating from our measured data. To keep the analysis concise, we assume that the nuclear spin magnetization in the $u$ coordinate space has a single period $\Omega$, i.e., the sample magnetization can be expressed as $p(u)g(u)$, where $g(u) = g(u + n\Omega)$, for all $n \in \mathbb{Z}$, and $p(u)$ is some slowly varying effective spin density of the sample. We remark that the assumption about the single period $\Omega$ is not essential and could be generalized rather easily. Eq. (1) in the main text identifies the in-phase and quadrature NMRd signals with the real and imaginary components of Fourier transformed $p(u)g(u)$, whereas the assumptions above imply that the Fourier transform of $p(u)g(u)$ is a train of diffraction echos, the envelopes of which are determined by the Fourier transform of $p(u)$. Moreover, because the square modulus of the Fourier transformed $p(u)$ is an even function, the squared sum of the in-phase and quadrature NMRd signals is a train of non-negative diffraction echo 'power peaks' that are symmetric around $n/\Omega$, $n \in \mathbb{Z}$.

Our analysis relies on slow variations in $p(u)$ for the detected NMRd power peaks to be unambiguously resolvable. This is clearly the case for the data shown in Fig. 4 of the main text. $\Omega$ can therefore be determined by evaluating the inverse centre for the first diffraction power peak:

$$\Omega = \frac{\sum_{t \in \mathcal{T}_{\mathrm{peak}}} \left[ E(S_t)^2 + E(Q_t)^2 \right]}{\sum_{t \in \mathcal{T}_{\mathrm{peak}}} t \left[ E(S_t)^2 + E(Q_t)^2 \right]} = f(\{x_t = E(S_t), y_t = E(Q_t)\}), \tag{S16}$$

where $\mathcal{T}_{\mathrm{peak}} \subseteq \mathcal{T}$ is the subset of measurements that lie in the region around the first diffraction echo in time and $f$ is defined as

$$
\begin{aligned}
f(\{x_t, y_t\}) &= \frac{\sum_{t \in \mathcal{T}_{\mathrm{peak}}} \left( x_t^2 + y_t^2 \right)}{\sum_{t \in \mathcal{T}_{\mathrm{peak}}} t \left( x_t^2 + y_t^2 \right)} \\
&= \Omega + \sum_{t \in \mathcal{T}_{\mathrm{peak}}} \alpha_t^{(S)} \left[ x_t - E(S_t) \right] + \sum_{t \in \mathcal{T}_{\mathrm{peak}}} \alpha_t^{(Q)} \left[ y_t - E(Q_t) \right] + \sum_{\substack{t_1 \in \mathcal{T}_{\mathrm{peak}} \\ t_2 \in \mathcal{T}_{\mathrm{peak}}}} \alpha_{t_1 t_2}^{(S,S)} \left[ x_{t_1} - E(S_{t_1}) \right] \left[ x_{t_2} - E(S_{t_2}) \right] \\
&\quad + 2 \sum_{\substack{t_1 \in \mathcal{T}_{\mathrm{peak}} \\ t_2 \in \mathcal{T}_{\mathrm{peak}}}} \alpha_{t_1 t_2}^{(S,Q)} \left[ x_{t_1} - E(S_{t_1}) \right] \left[ y_{t_2} - E(Q_{t_2}) \right] + \sum_{\substack{t_1 \in \mathcal{T}_{\mathrm{peak}} \\ t_2 \in \mathcal{T}_{\mathrm{peak}}}} \alpha_{t_1 t_2}^{(Q,Q)} \left[ y_{t_1} - E(Q_{t_1}) \right] \left[ y_{t_2} - E(Q_{t_2}) \right] + \dots \,.
\end{aligned}
\tag{S17}
$$

$\alpha, \alpha_t^{(S)}, \alpha_t^{(Q)}, \alpha_{t_1 t_2}^{(S,S)}, \alpha_{t_1 t_2}^{(S,Q)}, \alpha_{t_1 t_2}^{(Q,Q)}, \dots$ in Eq. (S17) above are the (normalized) Taylor coefficients in expanding $f$ around $\{x_t = E(S_t), y_t = E(Q_t)\}$. It is implicitly assumed with Eq. (S16) that the measurements contained in $\mathcal{T}_{\mathrm{peak}}$ sample the diffraction echo region sufficiently finely within its non-zero amplitude region so that the resulting error from finite sampling density is small compared to other error sources – a requirement that is easily fulfilled in practice.

Despite us not having access to $\{E(S_t), E(Q_t)\}$, it is reasonable to consider a random variable (an estimator [9])

$$\hat{\Omega} = f(\{S_t, Q_t\}) = \frac{\sum_{t \in \mathcal{T}_{\mathrm{peak}}} \left( S_t^2 + Q_t^2 \right)}{\sum_{t \in \mathcal{T}_{\mathrm{peak}}} t \left( S_t^2 + Q_t^2 \right)} \tag{S18}$$

for the purposes of estimating the grating period $\Omega$. It is clear that in general $E\left(\hat{\Omega}\right) \neq \Omega$ (implying that the estimator

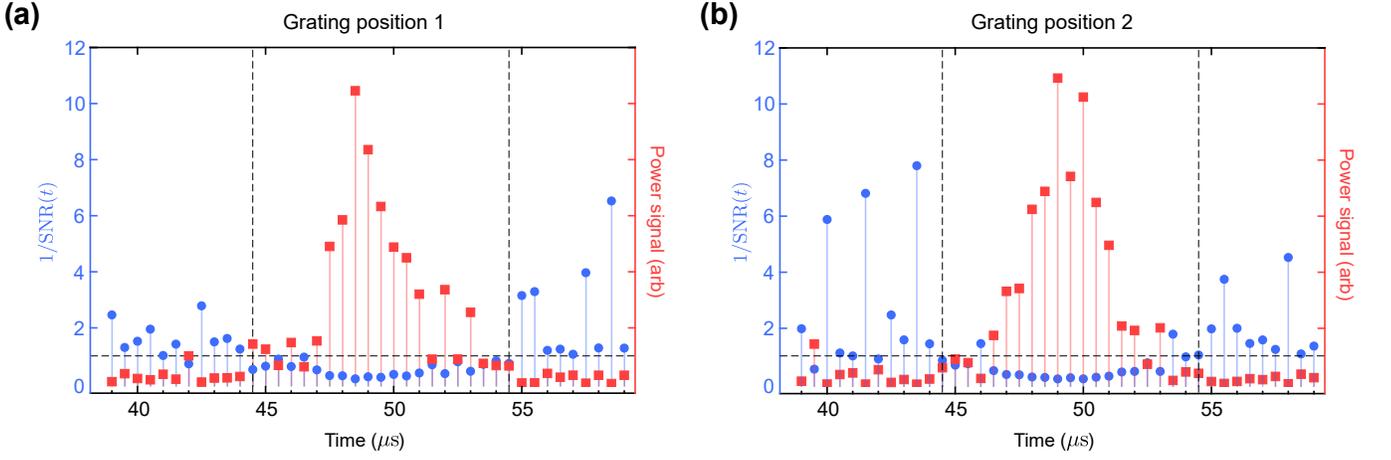

FIG. S12.    (a, b) NMRD 'power' signal $(s_t^2 + q_t^2)$ (red) for grating position 1 and 2, respectively.    Inverse SNR $\left[\sqrt{(2\sigma_t s_t)^2 + (2\varphi_t q_t)^2}/(s_t^2 + q_t^2)\right]$ (blue) for the same respective grating positions. The vertical lines mark the $t_{\min}$ and $t_{\max}$ values that are used for setting the interval of $\mathcal{T}_{\text{peak}}$ for the estimators of $\Omega$ and $\Delta u$, they exclude the data regions where the inverse SNR consistently exceeds the value 1.

is biased [9]), yet by verifying that our estimate for

$$
\begin{aligned}
\mathrm{E}\left(\hat{\Omega}\right) - \Omega &= \sum_{t \in \mathcal{T}_{\text{peak}}} \alpha_{tt}^{(S,S)} \mathrm{Var}\left(S_t\right) + \sum_{t \in \mathcal{T}_{\text{peak}}} \alpha_{tt}^{(Q,Q)} \mathrm{Var}\left(Q_t\right) + \mathcal{O}\left(\alpha_{ttt}\right) \\
&= \sum_{t_1 \in \mathcal{T}_{\text{peak}}} \frac{\left[\mathrm{E}\left(S_{t_1}\right)^2 \mathrm{Var}\left(S_{t_1}\right) + \mathrm{E}\left(Q_{t_1}\right)^2 \mathrm{Var}\left(Q_{t_1}\right)\right]\left(-4t_1\right)\sum_{t_2 \in \mathcal{T}_{\text{peak}}}\left(t_2 - t_1\right)\left[\mathrm{E}\left(S_{t_2}\right)^2 + \mathrm{E}\left(Q_{t_2}\right)^2\right]}{\left(\sum_{t_2 \in \mathcal{T}_{\text{peak}}} t_2\left[\mathrm{E}\left(S_{t_2}\right)^2 + \mathrm{E}\left(Q_{t_2}\right)^2\right]\right)^3} \\
&\quad + \sum_{t_1 \in \mathcal{T}_{\text{peak}}} \frac{\left[\mathrm{Var}\left(S_{t_1}\right) + \mathrm{Var}\left(Q_{t_1}\right)\right]\sum_{t_2 \in \mathcal{T}_{\text{peak}}}\left(t_2 - t_1\right)\left[\mathrm{E}\left(S_{t_2}\right)^2 + \mathrm{E}\left(Q_{t_2}\right)^2\right]}{\left(\sum_{t_2 \in \mathcal{T}_{\text{peak}}} t_2\left[\mathrm{E}\left(S_{t_2}\right)^2 + \mathrm{E}\left(Q_{t_2}\right)^2\right]\right)^2} + \mathcal{O}\left(\alpha_{ttt}\right)
\end{aligned}
\tag{S19}
$$

is negligible, we can safely use $\hat{\Omega}$ to determine the period (i.e. $\hat{\Omega}$ is as an approximately unbiased estimator). Along with the period of the grating, we also estimate the uncertainty in our estimate for $\Omega$. From the Taylor expansion in Eq. (S17) it follows that

$$
\begin{aligned}
\mathrm{Var}\left(\hat{\Omega}\right) = \mathrm{Var}\left(f\left[\{S_t, Q_t\}\right]\right) &= \sum_{t \in \mathcal{T}_{\text{peak}}}\left(\alpha_t^{(S)}\right)^2 \mathrm{Var}\left(S_t\right) + \sum_{t \in \mathcal{T}_{\text{peak}}}\left(\alpha_t^{(Q)}\right)^2 \mathrm{Var}\left(Q_t\right) \\
&\quad + 2\sum_{\substack{t_1 \in \mathcal{T}_{\text{peak}} \\ t_2 \in \mathcal{T}_{\text{peak}}}}\left(\alpha_{t_1 t_2}^{(S,S)}\right)^2 \mathrm{Var}\left(S_{t_1}\right)\mathrm{Var}\left(S_{t_2}\right) + 4\sum_{\substack{t_1 \in \mathcal{T}_{\text{peak}} \\ t_2 \in \mathcal{T}_{\text{peak}}}}\left(\alpha_{t_1 t_2}^{(S,Q)}\right)^2 \mathrm{Var}\left(S_{t_1}\right)\mathrm{Var}\left(Q_{t_2}\right) \\
&\quad + 2\sum_{\substack{t_1 \in \mathcal{T}_{\text{peak}} \\ t_2 \in \mathcal{T}_{\text{peak}}}}\left(\alpha_{t_1 t_2}^{(Q,Q)}\right)^2 \mathrm{Var}\left(Q_{t_1}\right)\mathrm{Var}\left(Q_{t_2}\right) + \mathcal{O}\left(\alpha_{ttt}\right) \\
&= 4\sum_{t_1 \in \mathcal{T}_{\text{peak}}} \frac{\left[\mathrm{E}\left(S_{t_1}\right)^2 \mathrm{Var}\left(S_{t_1}\right) + \mathrm{E}\left(Q_{t_1}\right)^2 \mathrm{Var}\left(Q_{t_1}\right)\right]\left(\sum_{t_2 \in \mathcal{T}_{\text{peak}}}\left(t_2 - t_1\right)\left[\mathrm{E}\left(S_{t_2}\right)^2 + \mathrm{E}\left(Q_{t_2}\right)^2\right]\right)^2}{\left(\sum_{t_2 \in \mathcal{T}_{\text{peak}}} t_2\left[\mathrm{E}\left(S_{t_2}\right)^2 + \mathrm{E}\left(Q_{t_2}\right)^2\right]\right)^4} + \mathcal{O}\left(\alpha_{tt}\right).
\end{aligned}
\tag{S20}
$$

To yield the most accurate and precise estimate for the grating period, we need to choose a region $\mathcal{T}_{\text{peak}}$ that attempts to minimize quantities in Eq. (S19, S20) while fully enclosing the first diffraction echo. Here, we regard $\mathcal{T}_{\text{peak}}$ as all data points within some interval $[t_{\min}, t_{\max}]$. To determine $t_{\min}$ and $t_{\max}$ in practice, we first estimate the 'signal-to-noise ratio' $\mathrm{SNR}(t) = \left(s_t^2 + q_t^2\right)/\sqrt{(2\sigma_t s_t)^2 + (2\varphi_t q_t)^2}$ for each $t \in \mathcal{T}$ in both grating position data sets. We then pick an interval $[t_{\min}, t_{\max}]$ as $\mathcal{T}_{\text{peak}}$ that excludes data regions where the inverse SNR consistently exceeds

the value 1. For both grating positions we choose $t_{\min} = 44.5~\mu s$ and $t_{\max} = 54.5~\mu s$ illustrated by Fig. S12 which also includes the diffraction echo 'power' signals calculated as $s_t^2 + q_t^2$. We have verified that our chosen $\mathcal{T}_{\text{peak}}$ sets are rather close – within a few end points included or excluded – to intervals that minimize our estimates for the quantity in Eq. (S19) (the bias of the estimator). We estimate the latter by substituting $\{ \mathrm{E}(S_t), \mathrm{E}(Q_t) \}$ with $\{ s_t, q_t \}$ and $\{ \mathrm{Var}(S_t), \mathrm{Var}(Q_t) \}$ with $\{ (\sigma_t)^2, (\varphi_t)^2 \}$, for all $t \in \mathcal{T}_{\text{peak}}$, in Eq. (S19) while ignoring terms of order $\alpha_{ttt}$ and higher.

Our estimates for $\Omega$ evaluate to 20.31 kHz and 20.27 kHz for grating positions 1 and 2, respectively, which are calculated by substituting $\{ \mathrm{E}(S_t), \mathrm{E}(Q_t) \}$ with $\{ s_t, q_t \}$ in Eq. (S19), where $\mathcal{T}_{\text{peak}}$ is taken to be the region shown in Fig. S12 and ignoring terms of order $\alpha_{ttt}$ and higher. Furthermore, we also estimate the uncertainty in the estimates above by approximating $\sqrt{\mathrm{Var}(\hat{\Omega})}$ in Eq. (S20) through the substitution of $\{ \mathrm{E}(S_t), \mathrm{E}(Q_t) \}$ with $\{ s_t, q_t \}$ and $\{ \mathrm{Var}(S_t), \mathrm{Var}(Q_t) \}$ with $\{ (\sigma_t)^2, (\varphi_t)^2 \}$, for all $t \in \mathcal{T}_{\text{peak}}$ and ignoring terms of order $\alpha_{tt}$ and higher. Hence, our final $\Omega$ estimates for grating positions 1 and 2 become $20.31 \pm 0.07$ kHz and $20.27 \pm 0.05$ kHz. We have also verified that the leading correction terms – $\mathcal{O}(\alpha_{tt})$ in Eq. (S20) – for our error estimates are more than an order of magnitude smaller than the quoted errors, and our estimates for $\mathrm{E}(\hat{\Omega}) - \Omega$ in Eq. (S19) are also at least an order of magnitude smaller than the quoted errors.

### B. Estimating the Grating Period Through Monte Carlo Sampling

In this subsection, we further verify the estimates of $\Omega$ in the previous subsection that were calculated by approximating $\mathrm{E}(\hat{\Omega})$ and $\mathrm{Var}(\hat{\Omega})$ in Eq. (S19) and Eq. (S20). We do this using Monte Carlo sampling. Rather than working with a fixed $\mathcal{T}_{\text{peak}}$ as before, we also attempt to account for the relative degree of ambiguity in picking $\mathcal{T}_{\text{peak}}$, as evident in Fig. S12. Accordingly, we take $t_{\min}$ and $t_{\max}$ to be realizations of two independent discrete random variables $X_{\min}$ and $X_{\max}$, the sample space of which is $\mathcal{T} = \{ 39~\mu s, 39.5~\mu s, 40~\mu s, ..., 58.5~\mu s, 59~\mu s \}$. Specifically, we use $X_{\min}$ and $X_{\max}$ that can be expressed as

$$X(t, \Delta t) = 1~\mu s \times \mathrm{Round}\left[ Y(t, \Delta t) \right]/2, \tag{S21}$$

where $Y(t, \Delta t)$ is a normally distributed random variable with an expectation value of $2t/(1~\mu s)$ and variance of $[2\Delta t/(1~\mu s)]^2$. $\Delta t$ here quantifies the uncertainty in either $t_{\min}$ or $t_{\max}$.

For the Monte Carlo sampling, we define a set of independent normally distributed random variables $\{ Z_t \} \cup \{ \Theta_t \}$, for $t \in \mathcal{T}$. We equate the expectation values and variances of $Z_t$ and $\Theta_t$ with our estimates for the expectation values and variances of $S_t$ and $Q_t$, such that $\mathrm{E}(Z_t) = s_t$, $\mathrm{E}(\Theta_t) = q_t$ and $\mathrm{Var}(Z_t) = (\sigma_t)^2$, $\mathrm{Var}(\Theta_t) = (\varphi_t)^2$. We slightly rewrite Eq. (S18) to account for the uncertainty in $\mathcal{T}_{\text{peak}}$:

$$\hat{\Omega}\left( \{ Z_t, \Theta_t \}; t_{\min}, t_{\max}, \Delta t \right) = \frac{\sum_{t \in \mathcal{T}_{\text{peak}}[X(t_{\min}, \Delta t), X(t_{\max}, \Delta t)]} \left( Z_t^2 + \Theta_t^2 \right)}{\sum_{t \in \mathcal{T}_{\text{peak}}[X(t_{\min}, \Delta t), X(t_{\max}, \Delta t)]} t \left( Z_t^2 + \Theta_t^2 \right)}, \tag{S22}$$

where $\mathcal{T}_{\text{peak}}[X(t_{\min}, \Delta t), X(t_{\max}, \Delta t)]$ is a shorthand for the collection of $t$ values within a random interval $[X(t_{\min}, \Delta t), X(t_{\max}, \Delta t)]$.

First, we verify the estimates and assertions in the preceding subsection by evaluating Eq. (S22) with $\Delta t \to 0$ after generating 100,000 (pseudo-)random instances of $\{ Z_t, \Theta_t \}$. For both grating positions, the resulting distribution of $\hat{\Omega}$ is extremely close to a Gaussian, with $\mathrm{E}\left[ \hat{\Omega}(t_{\min} = 44.5~\mu s, t_{\max} = 54.5~\mu s, \Delta t \to 0) \right] = 20.30$ kHz and $\sqrt{\mathrm{Var}\left[ \hat{\Omega}(44.5~\mu s, 54.5~\mu s, \Delta t \to 0) \right]} = 0.07$ kHz for position 1 and $\mathrm{E}\left[ \hat{\Omega}(44.5~\mu s, 54.5~\mu s, \Delta t \to 0) \right] = 20.26$ kHz and $\sqrt{\mathrm{Var}\left[ \hat{\Omega}(44.5~\mu s, 54.5~\mu s, \Delta t \to 0) \right]} = 0.05$ kHz for position 2. Finally, for both grating positions, we evaluate Eq. (S22) for 100,000 (pseudo-)random realizations of $\{ X(t_{\min}, \Delta t), X(t_{\max}, \Delta t) \} \cup \{ Z_t, \Theta_t \}$ for $\Delta t = 1.5~\mu s$, thereby introducing some variability to $\mathcal{T}_{\text{peak}}$. The resulting distributions of $\hat{\Omega}$ are reasonably closely approximated by Gaussians, and yield $\mathrm{E}\left[ \hat{\Omega}(t_{\min} = 44.5~\mu s, t_{\max} = 54.5~\mu s, \Delta t = 1.5~\mu s) \right] = 20.30$ kHz and $\sqrt{\mathrm{Var}\left[ \hat{\Omega}(44.5~\mu s, 54.5~\mu s, 1.5~\mu s) \right]} = 0.14$ kHz for position 1 and $\mathrm{E}\left[ \hat{\Omega}(44.5~\mu s, 54.5~\mu s, 1.5~\mu s) \right] = 20.26$ kHz and $\sqrt{\mathrm{Var}\left[ \hat{\Omega}(44.5~\mu s, 54.5~\mu s, 1.5~\mu s) \right]} = 0.10$ kHz for position 2. The estimates and errors quouted in the main text are the latter here.

## C. Estimating the Grating Translation

In this subsection, we estimate the translation $\Delta u$ by constructing a maximum likelihood estimator (MLE) [9] starting from Eq. (2) in the main text. Under the assumption of $p(u + \Delta u) \approx p(u)$, the expectation values of $\{S_t^p\}$ and $\{Q_t^g\}$, at positions $p \in \{\text{I}, \text{II}\}$, are related through a rotation matrix

$$\begin{bmatrix} \text{E}\left(S_t^\text{I}\right) \\ \text{E}\left(Q_t^\text{I}\right) \end{bmatrix} = \hat{R}(\Delta u, \tau) \begin{bmatrix} \text{E}\left(S_t^\text{II}\right) \\ \text{E}\left(Q_t^\text{II}\right) \end{bmatrix} = \begin{bmatrix} \cos(2\pi\Delta u\tau) & -\sin(2\pi\Delta u\tau) \\ \sin(2\pi\Delta u\tau) & \cos(2\pi\Delta u\tau) \end{bmatrix} \begin{bmatrix} \text{E}\left(S_t^\text{II}\right) \\ \text{E}\left(Q_t^\text{II}\right) \end{bmatrix}. \tag{S23}$$

For any $t \in \mathcal{T}$, let $\Upsilon_{1t}$ and $\Upsilon_{2t}$ be two random variables defined as

$$\Upsilon_{1t} = S_t^\text{II} \cos\left(2\pi\Delta ut\right) - Q_t^\text{II} \sin\left(2\pi\Delta ut\right) - S_t^\text{I},$$
$$\Upsilon_{2t} = Q_t^\text{II} \cos\left(2\pi\Delta ut\right) + S_t^\text{II} \sin\left(2\pi\Delta ut\right) - Q_t^\text{I}.$$

Because $\{S_t^p\}$ and $\{Q_t^p\}$ are independent normally distributed random variables, $\{\Upsilon_{1t}\}$ and $\{\Upsilon_{2t}\}$ are also independent and normally distributed as long as $\text{Var}\left(S_t^\text{II}\right) = \text{Var}\left(Q_t^\text{II}\right)$. Here, we assume these variances to be equal, which is justified given the identical spin detection protocol for the in-phase and quadrature measurements. In such case $\text{E}\left(\Upsilon_{1t}\right) = \text{E}\left(\Upsilon_{2t}\right) = 0$ and

$$\text{Var}\left(\Upsilon_{1t}\right) = \text{Var}\left(S_t^\text{II}\right) \cos^2\left(2\pi\Delta ut\right) + \text{Var}\left(Q_t^\text{II}\right) \sin^2\left(2\pi\Delta ut\right) + \text{Var}\left(S_t^\text{I}\right),$$
$$\text{Var}\left(\Upsilon_{2t}\right) = \text{Var}\left(Q_t^\text{II}\right) \cos^2\left(2\pi\Delta ut\right) + \text{Var}\left(S_t^\text{II}\right) \sin^2\left(2\pi\Delta ut\right) + \text{Var}\left(Q_t^\text{I}\right).$$

Let $\{z_t^p\}$ and $\{\theta_t^p\}$ be particular realizations of $\{S_t^p\}$ and $\{Q_t^p\}$, $p \in \{\text{I}, \text{II}\}$ and $t \in \mathcal{T}$. The probability density function (pdf) for $v_{1t} = z_t^\text{II} \cos\left(2\pi\Delta ut\right) - \theta_t^\text{II} \sin\left(2\pi\Delta ut\right) - z_t^\text{I}$ is then

$$f_{\Upsilon_{1t}}(\Upsilon_{1t} = v_{1t}; \Delta u) \propto \exp\left(-\frac{\left[z_t^\text{II} \cos\left(2\pi\Delta ut\right) - \theta_t^\text{II} \sin\left(2\pi\Delta ut\right) - z_t^\text{I}\right]^2}{2\left[\text{Var}\left(S_t^\text{II}\right) \cos^2\left(2\pi\Delta ut\right) + \text{Var}\left(Q_t^\text{II}\right) \sin^2\left(2\pi\Delta ut\right) + \text{Var}\left(S_t^\text{I}\right)\right]}\right),$$

while the pdf for $v_{2t} = \theta_t^\text{II} \cos\left(2\pi\Delta ut\right) + z_t^\text{II} \sin\left(2\pi\Delta ut\right) - \theta_t^\text{I}$ is

$$f_{\Upsilon_{2t}}(\Upsilon_{2t} = v_{2t}; \Delta u) \propto \exp\left(-\frac{\left[\theta_t^\text{II} \cos\left(2\pi\Delta ut\right) + z_t^\text{II} \sin\left(2\pi\Delta ut\right) - \theta_t^\text{I}\right]^2}{2\left[\text{Var}\left(Q_t^\text{II}\right) \cos^2\left(2\pi\Delta ut\right) + \text{Var}\left(S_t^\text{II}\right) \sin^2\left(2\pi\Delta ut\right) + \text{Var}\left(Q_t^\text{I}\right)\right]}\right).$$

Hence, given the independence of $\{\Upsilon_{it}\}$, for $i \in \{1, 2\}$ and $t \in \mathcal{T}$, we can combine the pdfs above into a likelihood function of $\Delta u$:

$$\mathcal{L}\left(\Delta u | \{\Upsilon_{1t} = v_{1t}, \Upsilon_{2t} = v_{2t}\}\right) \propto \prod_{t \in \mathcal{T}} \exp\left(-\frac{\left[z_t^\text{II} \cos\left(2\pi\Delta ut\right) - \theta_t^\text{II} \sin\left(2\pi\Delta ut\right) - z_t^\text{I}\right]^2}{2\left[\text{Var}\left(S_t^\text{II}\right) \cos^2\left(2\pi\Delta ut\right) + \text{Var}\left(Q_t^\text{II}\right) \sin^2\left(2\pi\Delta ut\right) + \text{Var}\left(S_t^\text{I}\right)\right]}\right) \times$$
$$\prod_{t \in \mathcal{T}} \exp\left(-\frac{\left[\theta_t^\text{II} \cos\left(2\pi\Delta ut\right) + z_t^\text{II} \sin\left(2\pi\Delta ut\right) - \theta_t^\text{I}\right]^2}{2\left[\text{Var}\left(Q_t^\text{II}\right) \cos^2\left(2\pi\Delta ut\right) + \text{Var}\left(S_t^\text{II}\right) \sin^2\left(2\pi\Delta ut\right) + \text{Var}\left(Q_t^\text{I}\right)\right]}\right).$$

A MLE $(\widehat{\Delta u})$ is then the value of $\Delta u$ that maximizes $\mathcal{L}$ for the data observed.

In practice we limit the data set employed for $\Delta u$ estimation, we use the same $\mathcal{T}_\text{peak} \subseteq \mathcal{T}$ as before, i.e., the region shown in Fig. S12. Furthermore, in place of the variances, $\{\text{Var}\left(S_t^p\right), \text{Var}(Q_t^p)\}$, we use our respective estimates $\left\{(\sigma_t^p)^2, (\varphi_t^p)^2\right\}$. Our MLE for the grating translation is then

$$\widehat{\Delta u}\left(\{S_t^\text{I}, Q_t^\text{I}, S_t^\text{II}, Q_t^\text{II}\}\right) = h\left(\{S_t^\text{I}, Q_t^\text{I}, S_t^\text{II}, Q_t^\text{II}\}\right), \tag{S24}$$

where

$$h\left(\{\alpha_t^\text{I}, \beta_t^\text{I}, \alpha_t^\text{II}, \beta_t^\text{II}\}\right) = \underset{\Delta u}{\arg\max} \left[\prod_{t \in \mathcal{T}_\text{peak}} \exp\left(-\frac{\left[\alpha_t^\text{II} \cos\left(2\pi\Delta ut\right) - \beta_t^\text{II} \sin\left(2\pi\Delta ut\right) - \alpha_t^\text{I}\right]^2}{2\left[\left(\sigma_t^\text{II}\right)^2 \cos^2\left(2\pi\Delta ut\right) + \left(\varphi_t^\text{II}\right)^2 \sin^2\left(2\pi\Delta ut\right) + \left(\sigma_t^\text{I}\right)^2\right]}\right) \times \right. \tag{S25}$$
$$\left. \prod_{t \in \mathcal{T}_\text{peak}} \exp\left(-\frac{\left[\beta_t^\text{II} \cos\left(2\pi\Delta ut\right) + \alpha_t^\text{II} \sin\left(2\pi\Delta ut\right) - \beta_t^\text{I}\right]^2}{2\left[\left(\varphi_t^\text{II}\right)^2 \cos^2\left(2\pi\Delta ut\right) + \left(\sigma_t^\text{I}\right)^2 \sin^2\left(2\pi\Delta ut\right) + \left(\varphi_t^\text{I}\right)^2\right]}\right)\right].$$

Similar to Eq. (S20) for the variance of $\hat{\Omega}$, we approximate the variance of the estimator $\widehat{\Delta u}$ [9] through Taylor expansion:

$$\mathrm{Var}\left(\widehat{\Delta u}\right) = \sum_{\substack{\tau \in \mathcal{T}_{\mathrm{peak}} \\ r \in \{\mathrm{I,II}\}}} \left(\frac{\partial h}{\partial \alpha_\tau^r}\bigg|_{\alpha_\tau^r = \mathrm{E}(S_\tau^r)}\right)^2 \mathrm{Var}\left(S_\tau^r\right) + \sum_{\substack{\tau \in \mathcal{T}_{\mathrm{peak}} \\ r \in \{\mathrm{I,II}\}}} \left(\frac{\partial h}{\partial \beta_\tau^r}\bigg|_{\beta_\tau^r = \mathrm{E}(Q_\tau^r)}\right)^2 \mathrm{Var}\left(Q_\tau^r\right) + \text{higher order terms}, \quad \text{(S26)}$$

where we have assumed the independence of $\{\Upsilon_{it}\}$ at different times. In order to estimate the translation $\Delta u$, we evaluate Eq. (S25) using numerical maximization in Mathematica [10], with $\mathcal{T}_{\mathrm{peak}} = \{44.5\ \mu s, 45\ \mu s, ..., 54.5\ \mu s\}$ and $\{\alpha_t^p = s_t^p, \ \beta_t^p = q_t^p\}$. Our corresponding estimate for $\Delta u$ is 4.67 kHz, while also estimate the uncertainty in this value by evaluating Eq. (S26) using finite differences for computing the partial derivatives of $h\left(\{\alpha_t^p, \beta_t^p\}\right)$, while substituting $\{\mathrm{E}(S_t^p), \mathrm{E}(Q_t^p)\}$ with $\{s_t^p, q_t^p\}$ and $\{\mathrm{Var}(S_t^p), \mathrm{Var}(Q_t^p)\}$ with $\{(\sigma_t^p)^2, (\varphi_t^p)^2\}$ and ignoring the higher order terms. The resulting estimate for $\sqrt{\mathrm{Var}\left(\widehat{\Delta u}\right)}$ is 0.20 kHz.

The estimator in Eq. (S24) is not explicitly invariant under swapping the data measured at position 1 with that of position 2, yet in the case of noiseless data the resulting estimate should only differ by a sign. We verify this by evaluating Eq. (S25) and (S26) exactly as it was described in the previous paragraph but with positions 1 and 2 swapped. This gives an estimate of $-4.69 \pm 0.19$ kHz for the translation $\Delta u$.

### D. Estimating the Grating Translation Through Monte Carlo Sampling

In this subsection, we verify our estimate for the translation and its uncertainty through Monte Carlo sampling. For that, we define a set of independent normally distributed random variables $\{Z_t^p, \Theta_t^p\}$ for $p \in \{\mathrm{I, II}\}$ and $t \in \mathcal{T}$. The expectation values and variances of these random variables are taken to be $\mathrm{E}(Z_t^p) = s_t^p$, $\mathrm{Var}(Z_t^p) = (\sigma_t^p)^2$ and $\mathrm{E}(\Theta_t^p) = q_t^p$, $\mathrm{Var}(\Theta_t^p) = (\varphi_t^p)^2$, i.e., our estimates for the means and variances of $\{S_t^p\}$ and $\{Q_t^p\}$, respectively. Similar to Sec. VI B, we allow for some uncertainty in $\mathcal{T}_{\mathrm{peak}}$ and use $\mathcal{T}_{\mathrm{peak}}\left[X(t_{\mathrm{min}}, \Delta t), X(t_{\mathrm{max}}, \Delta t)\right]$, with $X$ defined by Eq. (S21), to determine the set of measurements used in evaluating

$$\widehat{\Delta u}\left(\{Z_t^p = z_t^p, \Theta_t^p = \theta_t^p\}; t_{\mathrm{min}}, t_{\mathrm{max}}, \Delta t\right) = \tag{S27}$$

$$\arg\max_{\Delta u} \left[\prod_{t \in \mathcal{T}_{\mathrm{peak}}[X(t_{\mathrm{min}}, \Delta t), X(t_{\mathrm{max}}, \Delta t)]} \exp\left(-\frac{\left[z_t^{\mathrm{II}} \cos\left(2\pi \Delta u t\right) - \theta_t^{\mathrm{II}} \sin\left(2\pi \Delta u t\right) - z_t^{\mathrm{I}}\right]^2}{2\left[(\sigma_t^{\mathrm{II}})^2 \cos^2\left(2\pi \Delta u t\right) + (\varphi_t^{\mathrm{II}})^2 \sin^2\left(2\pi \Delta u t\right) + (\sigma_t^{\mathrm{I}})^2\right]}\right) \times$$

$$\prod_{t \in \mathcal{T}_{\mathrm{peak}}[X(t_{\mathrm{min}}, \Delta t), X(t_{\mathrm{max}}, \Delta t)]} \exp\left(-\frac{\left[\theta_t^{\mathrm{II}} \cos\left(2\pi \Delta u t\right) + z_t^{\mathrm{II}} \sin\left(2\pi \Delta u t\right) - \theta_t^{\mathrm{I}}\right]^2}{2\left[(\varphi_t^{\mathrm{II}})^2 \cos^2\left(2\pi \Delta u t\right) + (\sigma_t^{\mathrm{II}})^2 \sin^2\left(2\pi \Delta u t\right) + (\varphi_t^{\mathrm{I}})^2\right]}\right)\right].$$

We calculate $\mathrm{E}\left[\widehat{\Delta u}\left(t_{\mathrm{min}} = 44.5\ \mu s, t_{\mathrm{max}} = 54.5\ \mu s, \Delta t \to 0\right)\right]$ and $\mathrm{Var}\left[\widehat{\Delta u}\left(t_{\mathrm{min}} = 44.5\ \mu s, t_{\mathrm{max}} = 54.5\ \mu s, \Delta t \to 0\right)\right]$ by evaluating Eq. (S27) for 10,000 (pseudo-)random realizations of $\{Z_t^p, \Theta_t^p\}$. As expected for a MLE estimator, the Monte-Carlo sampled $\mathrm{E}\left(\widehat{\Delta u}\right)$ and $\sqrt{\mathrm{Var}\left(\widehat{\Delta u}\right)}$ perfectly confirm the estimates quoted in the previous subsection, while the distribution of $\widehat{\Delta u}$ is well-described by a Gaussian. We also compute 10,000 instances of $\widehat{\Delta u}\left(t_{\mathrm{min}} = 44.5\ \mu s, t_{\mathrm{max}} = 54.5\ \mu s, \Delta t = 1.5\ \mu s\right)$ in (S27) by generating 10,000 (pseudo-)random realizations of $\{X(t_{\mathrm{min}}, \Delta t), \ X(t_{\mathrm{max}}, \Delta t)\} \cup \{Z_t^p, \Theta_t^p\}$. The resulting distribution of $\widehat{\Delta u}$ is again very close to a Gaussian distribution and yields $\widehat{\Delta u} = 4.69$ kHz and $\sqrt{\mathrm{Var}(\widehat{\Delta u})} = 0.20$ kHz further confirming the MLE estimates in the previous subsection.

## VII. PIEZO TRANSDUCER CHARACTERIZATION

The interferometric displacement detection measurements presented in Sec. V of the main text were made using a lead zirconate titanate piezoceramic EBL#3 purchased from EBL Piezoelectric Precision. The piezoelectric transducer (PT) was cut into a $1.5\,\mathrm{mm} \times 1.0\,\mathrm{mm} \times 0.31\,\mathrm{mm}$ piece and glued to the silicon substrate containing the SiNW oscillator [Fig. S13(a)]. The PT was poled in the shear direction, which was aligned along the $y$ axis. We characterized the

room temperature response of the assembly shown in Fig. S13(a) by using an optical interferometer to measure the displacement in the $y$ and $z$ directions for a given voltage $V_P$ applied to the PT. These measurements were done with a different interferometer setup than the one shown in the figure, which allowed us to measure the displacement in both the $y$ and $z$ directions. The setup used for the experiment only permits interferometric detection in the $y$ direction.

The assembly used for the room temperature tests was identical in construction to the one used in the measurements. For $V_P = 7.2$ V, $\delta y = 56$ Å and $\delta z = 2.2$ Å. The measured value of $\delta y/V_P = 7.8$ Å/V compares very well with the tabulated value of the shear displacement constant $d_{15} = 7.3$ Å/V for the EBL#3 piezoceramic at room temperature.

To determine the response time of the PT, we use the optical interferometer to measure the $y$-axis displacement of the Si chip in response to a voltage pulse applied to the PT. The data in Fig. S13(b) shows data measured at 4 K in the same configuration used to take the data presented in Sec. V of the main text. At 4 K for $V_P = 7.2$ V, $\delta y = 10$ Å which is a factor of 5.6 smaller than the room temperature value. Likewise, the $z$-axis displacement corresponding to $V_P = 7.2$ V at 4 K ($\delta z = 0.81$ Å), determined from the fit to the data in Fig.6(c), is a factor of 2.7 smaller than the measured value at room temperature. The reduction in the piezo coefficient at low temperature is expected; the amount of reduction depends on the details of the piezo material and geometry.

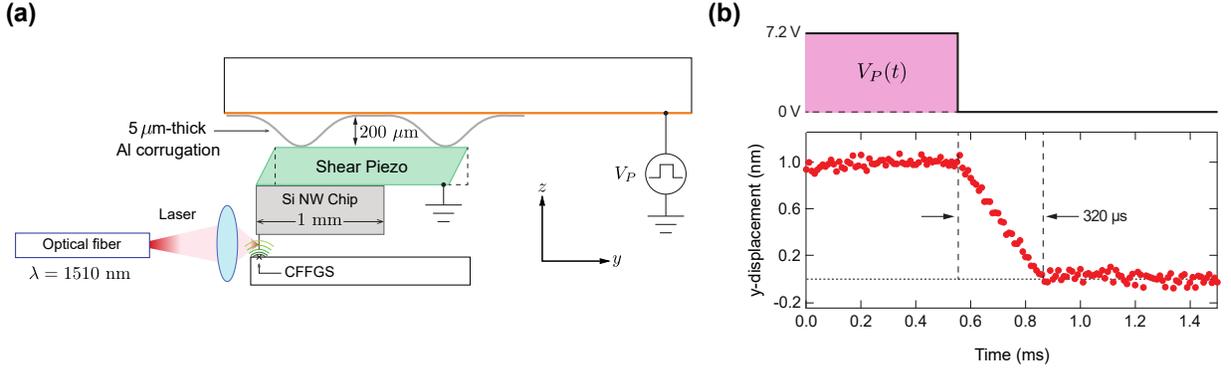

FIG. S13. (a) Schematic of the experimental setup showing the configuration of the optical interferometer used for displacement detection of the SiNW oscillator, the CFFGS, and the SiNW chip mounted on the shear PT. The PT assembly is glued to a 5-$\mu$m thick Al corrugation, which serves to isolate the mechanical modes of the PT assembly from rest of the mechanical assembly. (b) Data showing the $y$-axis displacement of the PT in response to a voltage pulse at 4 K, measured using the optical interferometer.

## VIII. NMR INTERFEROMETRY DATA ANALYSIS

Here, we provide details on the fits conducted for extracting the sample displacement from the NMR interferometry data (Sec. V of the main text).

### A. Fit using Sample Geometry

Our main method for extracting the displacement from both the $\tau_e = \tau_d = 960$ $\mu$s and $\tau_e = 480$ $\mu$s data is a least-squares fit using the InP sample geometry characterized in Sec. III. The fit function is constructed using Eq. (3, A1) of the main text:

$$\begin{bmatrix} s_I(\tau_d) \\ s_Q(\tau_d) \end{bmatrix} = \frac{A}{s_0} \sum_{\mathbf{r}} G_R^2(\mathbf{r}) \chi(u(\mathbf{r})) \begin{bmatrix} \cos\varphi(\mathbf{r}) \\ \sin\varphi(\mathbf{r}) \end{bmatrix},$$ (S28)

where $A$ is an unknown proportionality factor, $s_0 = \sum_{\mathbf{r}} G_R^2(\mathbf{r})\chi(u(\mathbf{r}))$ is a normalization, $\varphi(\mathbf{r}) = 2\pi\left[\tau_e u(\mathbf{r}) - \tau_d u(\mathbf{r} + \delta\mathbf{r})\right]$ and the sum runs over ~25,000 points sampled uniformly from the InP NW geometry. The Rabi frequency $u(\mathbf{r})$ and readout gradient $G_R(\mathbf{r})$ are calculated using the CFFGS field simulations of Sec. II. The $\chi(u)$ filter function is [6]

$$\chi(u) = |f_{\mathrm{AFP}}(u)|^N \left(\frac{1}{N+1} \frac{1 - |f_{\mathrm{AFP}}(u)|^{N+1}}{1 - |f_{\mathrm{AFP}}(u)|}\right)^2 \theta(650 \text{ kHz} - u),$$ (S29)

where $\theta$ is the Heaviside step function, $f_{AFP}(u) = 1 - 2|\langle\downarrow|U_{AFP}(u)|\uparrow\rangle|^2$, and $U_{AFP}(u)$ is the AFP propagator for a spin experiencing Rabi frequency $u$, while $N \approx 14,000$ denotes the number of AFPs applied within a measurement block. The step function represents the effect of the 650 kHz low-pass filter, described in Sec. IV B., in the beginning of the NMR encoding block.

The fit to the $\tau_e = \tau_d = 960$ $\mu$s data is done by assuming $\delta\mathbf{r} = \alpha V_P(\hat{\mathbf{z}} + \beta\hat{\mathbf{y}})$, where $\alpha$ (the piezo coefficient), $\beta$ (the $\delta y/\delta z$ ratio) and the overall signal amplitude are taken to be the fit parameters. Table I depicts the resulting optimal parameters along with the corresponding covariances [11].

| Parameter | Best fit |
|-----------|----------|
| $A$ | 0.97 |
| $\beta$ | 6.64 |
| $\alpha$ (Å/V) | 0.15 |

| Parameter | $A$ | $\beta$ | $\alpha$ |
|-----------|-----|---------|----------|
| $A$ | 0.002 | 0.0045 | $7 \times 10^{-5}$ Å/V |
| $\beta$ | 0.0045 | 0.16 | $-6 \times 10^{-4}$ Å/V |
| $\alpha$ | $7 \times 10^{-5}$ Å/V | $-6 \times 10^{-4}$ Å/V | $3 \times 10^{-5}$ (Å/V)$^2$ |

TABLE I. Details of the fit to the $\tau_e = \tau_d = 960$ $\mu$s interferometry data. (Left) Optimal fit parameters (Right) Covariance matrix elements of the fit. Each cell indicates the covariance of the parameters in the corresponding row and column.

We thus determine the piezo coefficient to be $\alpha = \delta z/V_P = 0.15 \pm 0.01$ Å/V, where the uncertainty is calculated using the corresponding diagonal element of the covariance matrix.

For the phase-sensitive $\tau_e = 480$ $\mu$s data sets, we similarly use Eq. (S28) with $\delta\mathbf{r} = 0$ for the $V_P = 0$ data, and $\delta\mathbf{r} = \delta y\ \hat{\mathbf{y}} + \delta z\ \hat{\mathbf{z}}$ for the $V_P = 7.2$ V data. The fit parameters are chosen to be $A$, $\delta y$ and $\delta z$, where $A$ is shared between data sets. We provide the best fit parameters and their covariances in Table II.

| Parameter | Best fit |
|-----------|----------|
| $A$ | 1.09 |
| $\delta z$ (Å) | 0.81 |
| $\delta y$ (Å) | 9.72 |

| Parameter | $A$ | $\delta z$ | $\delta y$ |
|-----------|-----|------------|------------|
| $A$ | 0.003 | 0.04 Å | -0.0003 Å |
| $\delta z$ | 0.04 Å | 1.80 Å$^2$ | -0.015 Å$^2$ |
| $\delta y$ | -0.0003 Å | -0.015 Å$^2$ | 0.005 Å$^2$ |

TABLE II. Details of the fit to the $\tau_e = 480$ $\mu$s phase-sensitive data. (Left) Optimal fit parameters (Right) Covariance matrix elements of the fit. Each cell indicates the covariance of the parameters in the corresponding row and column.

The longitudinal displacement of the sample is thus calculated to be $\delta z = 0.81 \pm 0.07$ Å. From the covariance matrix, we also recognize that the correlation coefficient of $\delta z$ and $\delta y$ is $\mathrm{Cov}(\delta z, \delta y)/\sqrt{\mathrm{Var}(\delta z)\mathrm{Var}(\delta y)} = -0.16$, indicating that the lateral and longitudinal displacements are only weakly correlated. This can also be seen by using Eq. (S28) to simulate the signal dependence on the two displacement directions. The results (Fig. S14) show that $\delta z$ controls the phase of the signal without substantially affecting the overall amplitude. On the other hand, the effect of a nanometer-scale $\delta y$ is to suppress the amplitude while leaving the phase unaffected. The two displacements thus affect the signal roughly independently. This observation is used in the next section to devise an alternative fitting procedure that is independent of our model for the sample geometry.

### B. Fit using Measured Rabi-Frequency Distribution

In light of the previous discussion, we construct an alternative fit function for the phase-sensitive $\tau_e = 480$ $\mu$s data that directly accounts for only the $\delta z$ displacement, while modeling the effect of $\delta y$ as a suppression in overall amplitude. This allows for constructing a fit model using the measured Rabi-frequency distribution $p(u)$ rather than the derived sample geometry. For a spin at position $\mathbf{r}$, the accrued phase is taken to be $\varphi(z) = 2\pi[\tau_e u(z) - \tau_d u(z + \delta z)] \approx 2\pi(\tau_e - \tau_d)u(z) - 2\pi\tau_d\delta z\ u'(z)$, since $u$ and $z$ have an approximate one-to-one relation in the absence of an explicit lateral displacement. The Taylor expansion $u(z + \delta z) \approx u(z) + \delta z\ u'(z)$ is justified as the Rabi frequency varies on a much larger length scale than a few Ångstroms. Therefore, the phase is identical for spins experiencing the same Rabi frequency, meaning that we can write the signal as an integral over the Rabi-frequency distribution. We thus use

$$\begin{bmatrix} s_I(\tau_d) \\ s_Q(\tau_d) \end{bmatrix} = A_1 \int_0^\infty du\ p(u) \begin{bmatrix} \cos\left[2\pi(\tau_e - \tau_d)u\right] \\ \sin\left[2\pi(\tau_e - \tau_d)u\right] \end{bmatrix}, \tag{S30}$$

to fit to the $V_P = 0$ data, and

$$\begin{bmatrix} s_I(\tau_d) \\ s_Q(\tau_d) \end{bmatrix} = A_2 \int_0^\infty du\ p(u) \begin{bmatrix} \cos\left[2\pi(\tau_e - \tau_d)u - 2\pi\tau_d\delta z\ u'(z(u))\right] \\ \sin\left[2\pi(\tau_e - \tau_d)u - 2\pi\tau_d\delta z\ u'(z(u))\right] \end{bmatrix}, \tag{S31}$$

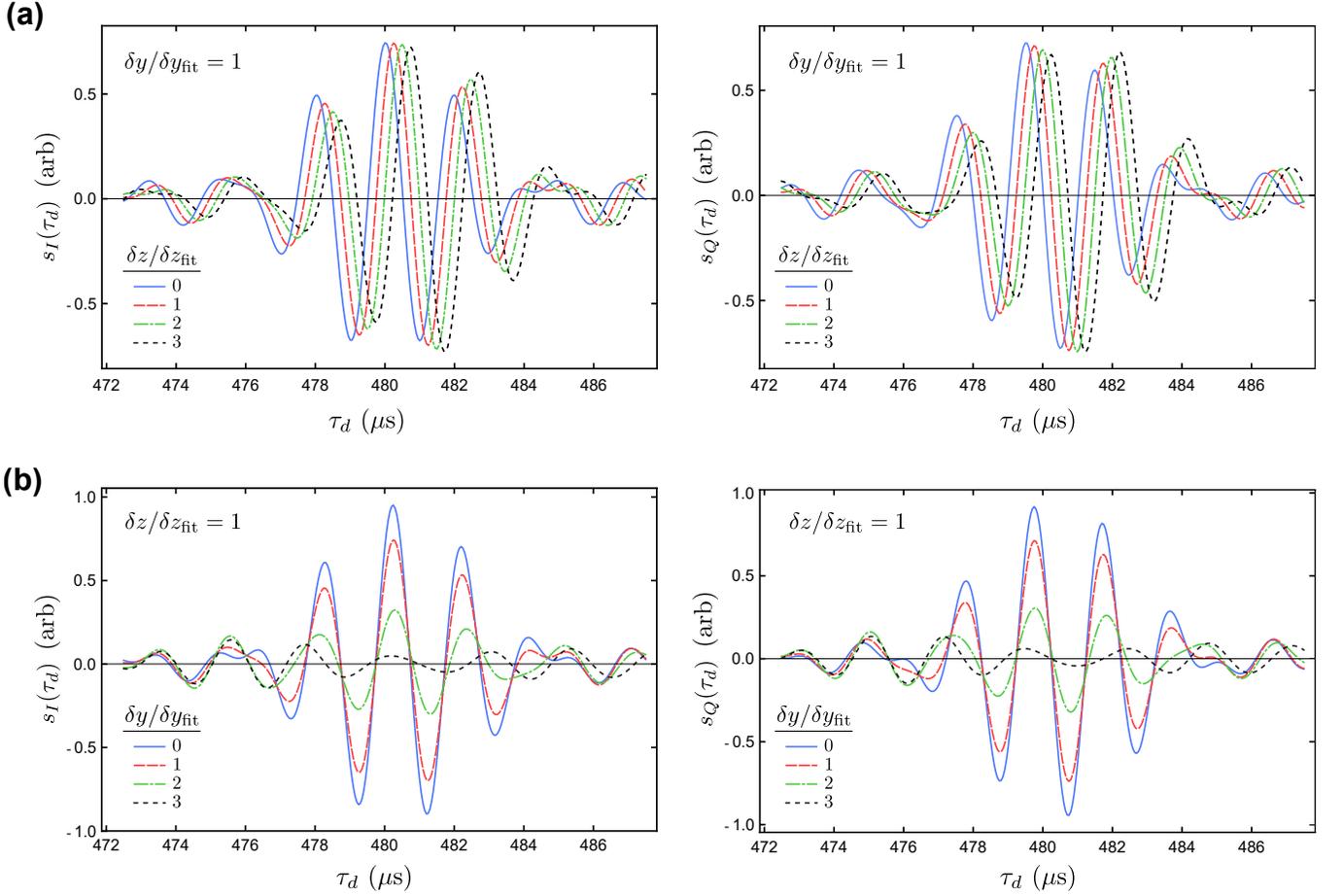

FIG. S14. Simulated signal dependence on the lateral and longitudinal displacements for the phase-sensitive $\tau_e = 480$ $\mu$s experiment. (a) In-phase (left) and quadrature (right) signals for different $\delta z$ values at constant $\delta y = \delta y_{\mathrm{fit}} = 9.72$ Å. (b) In-phase (left) and quadrature (right) signals for different $\delta y$ values at constant $\delta z = \delta z_{\mathrm{fit}} = 0.81$ Å.

to fit to the $V_P = 7.2$ V data, with $A_1$, $A_2$ and $\delta z$ being fit parameters. We used the measured $p(u)$ of Fig. 6(a) in the main text, which is interpolated for calculating the integral. We normalize $p(u)$ such that $\int du\ p(u) = 1$, while setting all values outside the [400 kHz, 650 kHz] window to zero. The resulting optimal parameters and the associated covariance matrix are presented in Table III, indicating that consistent with Sec. VIII A, $\delta z = 0.85 \pm 0.07$ Å. The fitted curve is depicted in Fig. S15.

| Parameter | Best fit |
|---|---|
| $A_1$ | 1.03 |
| $A_2$ | 0.83 |
| $\delta z$ (Å) | 0.85 |

| Parameter | $A_1$ | $A_2$ | $\delta z$ |
|---|---|---|---|
| $A_1$ | 0.003 | 0 | 0 |
| $A_2$ | 0 | 0.003 | $2 \times 10^{-6}$ Å |
| $\delta z$ | 0 | $2 \times 10^{-6}$ Å | 0.005 Å$^2$ |

TABLE III. Details of the alternative fitting method for the phase-sensitive $\tau_e = 480$ $\mu$s data. (Left) The optimal fit parameters. (Right) Covariance matrix elements of the fit parameters.

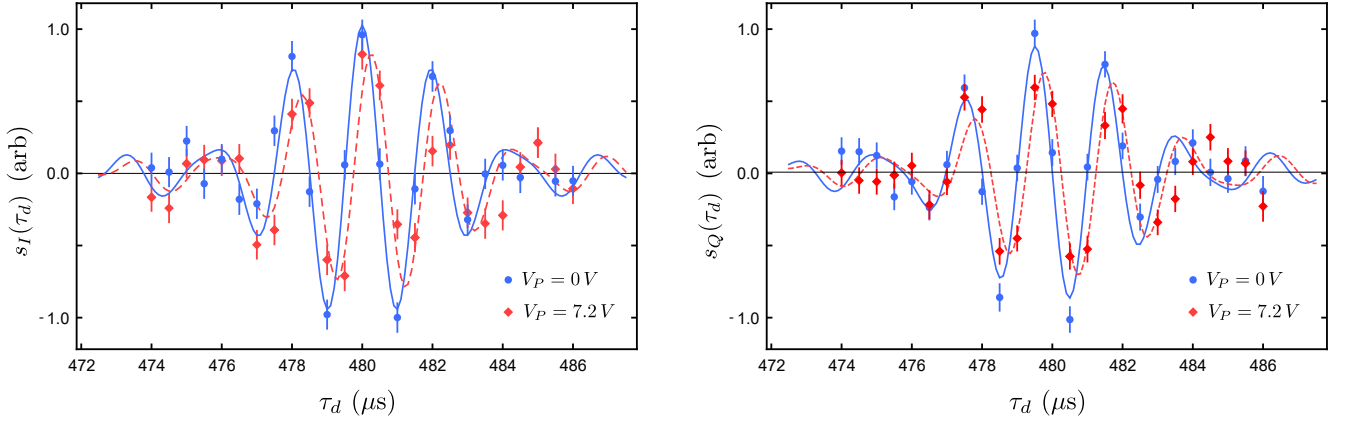

FIG. S15. Alternative fit to the phase-sensitive 480 $\mu$s data based on the measured Rabi-frequency distribution.


\end{document}